\documentclass{aa}
\usepackage{txfonts}
\usepackage{natbib}
\usepackage{graphicx}
\usepackage[english]{babel}
\usepackage{lscape}
\usepackage{multirow}
\usepackage{hyperref}

\newcommand{\rbracket}{]}

\begin{document}

   \title{Effects of Helium Enrichment in Globular Clusters I.Theoretical Plane with PGPUC stellar evolution code}

   \author{A.~A.~R. Valcarce$^{1,2,3,4}$, M. Catelan$^{1,3,4}$ \and A. V. Sweigart$^{5}$}
   \offprints{A.~A.~R. Valcarce}
   \institute{
	      $^1$Pontificia Universidad Cat\'olica de Chile, Facultad de F\'isica, Departamento de Astronom\'ia y Astrof\'isica,
              Av. Vicu\~na Mackena 4860, 782-0436 Macul, Santiago, Chile\\ \email{avalcarc@astro.puc.cl; mcatelan@astro.puc.cl}\\
              $^2$Universidade Federal do Rio Grande do Norte, Departamento de F\'isica, 59072-970 Natal, RN, Brazil\\ \email{avalcarc@dfte.ufrn.cl} \\
	      $^3$The Milky Way Millennium Nucleus, Av. Vicu\~{n}a Mackenna 4860, 782-0436, Macul, Santiago, Chile \\
	      $^4$Pontificia Universidad Cat\'olica de Chile, Centro de Astroingenier\'ia, Av. Vicu\~na Mackena 4860, 782-0436 Macul, Santiago, Chile\\ 
	      $^5$NASA Goddard Space Flight Center, Exploration of the Universe Division, Code 667, Greenbelt, MD 20771, USA\\ \email{allen.v.sweigart@nasa.gov}
             }
   \date{Received 30 April 2012 / Accepted 23 August 2012}

  \abstract
   {}
   {Recently, the study of globular cluster (GC) color-magnitude diagrams (CMDs) has shown that some of them harbor multiple populations with different chemical compositions and/or ages. In the first case, the most common candidate is a spread in the initial helium abundance, but this quantity is difficult to determine spectroscopically due to the fact that helium absorption lines are not present in cooler stars, whereas for hotter GC stars gravitational settling of helium becomes important. As a consequence, indirect methods to determine the initial helium abundance among populations are necessary. For that reason, in this series of papers, we investigate the effects of a helium enrichment in populations covering the range of GC metallicities.
   }
   {In this first paper, we present the theoretical evolutionary tracks, isochrones, and zero-age horizontal branch (ZAHB) loci calculated with the Princeton-Goddard-PUC (PGPUC) stellar evolutionary code, which has been updated with the most recent input physics and compared with other theoretical databases. The chemical composition grid covers 9 metallicities ranging from Z=1.60$\times 10^{-4}$ to 1.57$\times 10^{-2}$ (-2.25$\lesssim$[Fe/H]$\lesssim$-0.25), 7 helium abundances from Y=0.230 to 0.370, and an alpha-element enhancement of [$\alpha$/Fe]=0.3.
   }
   {The effects of different helium abundances that can be observed in isochrones are: splits in the main sequence (MS), differences in the luminosity ($L$) and effective temperature ($T_{\rm eff}$) of the turn off point, splits in the sub giant branch being more prominent for lower ages or higher metallicities, splits in the lower red giant branch (RGB) being more prominent for higher ages or higher metallicities, differences in $L$ of the RGB bump (with small changes in $T_{\rm eff}$), and differences in $L$ at the RGB tip. At the ZAHB, when Y is increased there is an increase (decrease) of $L$ for low (high) $T_{\rm eff}$, which is affected in different degrees depending on the age of the GC being studied. Finally, the ZAHB morphology distribution depending on the age explains how for higher GC metallicities a population with higher helium abundance could be hidden at the red ZAHB locus.
   }
   {}


   \authorrunning{Valcarce, Catelan, \& Sweigart}
   \titlerunning{Effects of He-Enrichment in GCs - I.Theoretical Plane with PGPUC}
   \maketitle
%

\section{Introduction}
\label{intro}

Globular clusters (GCs) are objects orbiting around the Milky Way and other galaxies and containing between $\sim$10$^4$ and $\sim$10$^7$ stars. Nowadays, $\sim$160 GCs are known in the Milky Way, with metallicities $[M/H] = \log (Z/X) -\log (Z/X)_\odot$ between $\sim$-2.3 and $\sim$-0.25 \citep[][revision 2010]{Harris1996} and with ages ranging from $\sim$ 6.5 to $\sim$13 Gyr \citep{Aparicio_etal2001, Chaboyer2001, MarinFranch_etal2009, Dotter_etal2010, MoniBidin_etal2011}. 

For several decades these objects have been considered as an excellent laboratory for the study of simple stellar populations (SSPs) because it was believed that all stars inside them were formed from a homogeneous cloud at the same time. However, it was recognized early that their horizontal branch (HB) morphologies could not be simply parametrized.  While the metallicity of a GC is the first parameter defining its HB morphology \citep{Sandage_Wallerstein1960}, empirical findings show clearly that it cannot be the only parameter at work. In fact, for many years we have been well aware of several GCs characterized by quite different HB morphologies in spite of their quite similar metallicity. For this reason, several {\em second parameters} have been invoked to solve this conundrum, where, of course, the most direct, and also the first invoked, second parameter was the age \citep{Sandage_Wallerstein1960, Rood_Iben1968}. This problem has been widely debated until the present date, but without reaching a consensus \citep[e.g.][ and references therein]{Catelan_etal1993, Lee_etal1994, Stetson_etal1996, Sarajedini_etal1997, Ferraro_etal1997c, Catelan2009b, Dotter_etal2010, Gratton_etal2010}.

Another early invoked candidate was the initial helium abundance \citep{van_den_Bergh1965, van_den_Bergh1967, Sandage_Wildey1967, Hartwick1968}, but the major difficulty with this element is its extremely difficult measurement, which requires high temperatures in order to be detected in stars. Moreover, blue HB stars hotter than $\sim$11\,500 K as well as stars on the so-called extreme HB suffer from metal levitation and helium settling \citep[e.g.,][]{Grundahl_etal1999,Behr2003}, leaving only a narrow band between $\sim$11\,500 and $\sim$8\,000 K for detecting the initial helium abundance in HB stars. Even so, its measurement remains difficult \citep[see][]{Villanova_etal2009, Villanova_etal2012}. Even though helium was one of the first candidates, this explanation of the second-parameter effect became less popular after the Milky Way formation scenario of \citet{Searle_Zinn1978}, where the age is assumed to be the second parameter. In addition, differences in helium among GCs also seemed less likely, due to the increasing evidence that GCs have similar helium abundances \citep[e.g.,][]{Salaris_etal2004}.

However, after the observational evidence of multiple main sequences (MSs) in GCs \citep{Norris2004, Piotto_etal2005, DAntona_etal2005, Piotto_etal2007}, the hypothesis that helium is the second parameter has regained strength, but in this case the difference in the initial helium abundance is mostly among stars inside the same GC \citep[][among several others]{DAntona_etal2002, DAntona_etal2005, Norris2004, Lee_etal2005, Piotto_etal2005, Piotto_etal2007}. In fact, it seems that the combination of age as the second parameter and a helium spread among stars as the ``third parameter'' is a promising attempt \citep{Gratton_etal2010}. 

It is worth noting that, although the existence of a correlation between the chemical abundance anomalies (such as light-element anticorrelations) and the HB morphology has been suggested for a long time \citep{Norris1981, Smith_Norris1983, Catelan_deFreitasPacheco1995}, only recently has direct evidence of this correlation been obtained, as in the case of the GC M4 \citep[][]{Marino_etal2011, Villanova_etal2012}.

As can be noted, the hypothesis that GCs are SSPs is being ruled out, due to the chemical inhomogeneities among stars of the same GC that have been (directly or indirectly) detected and associated to different star formation episodes in the same GC \citep[][and references therein]{Decressin_etal2007b, DErcole_etal2008, Carretta_etal2010, Conroy_Spergel2011, Bekki2011, Valcarce_Catelan2011}.

In addition to the presence of multiple populations in GCs, another development makes the availability of evolutionary tracks for different helium abundances highly necessary: as pointed out by \citet{Catelan_deFreitasPacheco1996} and \citet{Catelan2009}, and also emphasized recently by \citet{Nataf_Gould2012}, \citet{Nataf_Udalski2011}, and \citet{Nataf_etal2011,Nataf_etal2011a}, stellar populations with different enrichment histories may have different helium enrichment ``laws" (i.e., different Y-Z relations).

In this article, we show the effects of the helium enrichment in isochrones and zero-age HB (ZAHB) loci in the theoretical Hertzsprung-Russell (HR) diagram. First in section \ref{PGPUCsec}, we present the Princeton-Goddard-PUC (PGPUC) stellar evolution code (SEC) which we used to create a set of evolutionary tracks, isochrones and ZAHB loci for different metallicities and helium abundances (section \ref{PGPUCdatabase}). Then, we show the effects of the helium enrichment (section \ref{EffectsHe}) and end with the conclusions (section \ref{conclu})\footnote{All of the models described in this paper can be downloaded from the ``PGPUC Online" dedicated website, as described in Appendix \ref{AppPGPUConline}.}.


\section{PGPUC stellar evolution code}
\label{PGPUCsec}

The PGPUC stellar evolution code (SEC) is an updated version of the code created by M. Schwarzschild and R. H{\"a}rm \citep{Schwarzschild_Harm1965, Harm_Schwarzschild1966}, and then highly modified by A. V. Sweigart \citep[][ and references therein]{Sweigart1971, Sweigart1973, Sweigart1994, Sweigart1997, Sweigart_Demarque1972, Sweigart_Gross1974, Sweigart_Gross1976, Sweigart_Catelan1998}. It is focused mainly on the study of low-mass stars, defined as stars that develop a degenerate helium core before the onset of helium burning. Since almost all the stars living in present-day GCs can be considered as low-mass stars, the PGPUC SEC is ideal to be used as the basis to create isochrones and ZAHBs for studying the GC properties.

In the following subsections we first explain some of the updates to the input physics included in the PGPUC SEC and then provide a comparison with other codes. Unless otherwise noted, all physical quantities are given in cgs units.

\subsection{Opacities}

For high-temperature opacities the original version of our SEC (hereafter called PG) used the Rosseland mean opacity tables calculated by the OPAL team \citep{Iglesias_Rogers_Wilson1987}, as distributed in 1992 \citep[][OPAL92]{Rogers_Iglesias1992}, where twelve metals were considered. In PGPUC, we used the most recent version of the OPAL tables \citep[][OPAL96]{Iglesias_Rogers1996}, where seven new metals are considered, and some errors in the original OPAL code were fixed.

In the PG SEC, conductive opacity was calculated using the \citet[][for $\log \rho <$ 6]{Hubbard_Lampe1969} and \citet[][for $\log \rho \ge $ 6]{Canuto1970} tables. However, these tables cover a limited mixture of elements, and their approximations are not completely valid when the matter is in a state between gas and liquid, which happens in the interior of low-mass stars in the RGB phase \citep{Catelan_etal1996,Catelan2009}. In the PGPUC SEC, we used the \citet[][hereafter C07]{Cassisi_etal2007} calculations for the conductive opacities, which cover the partial and high electron degeneracy regimes. These calculations include electron-ion, electron-phonon, and electron-electron scattering.

When the temperature drops below $\log T\sim$ 4.5, opacity sources that are not considered in the high-temperature regime begin to be important. These sources include: i) negative H ion, which is a large source of continuum opacity; ii) atoms that are not completely ionized; iii) molecules that are formed at low temperatures; and iv) dust grains, which are important at very low temperatures. These opacity sources are important for the outer layers of cool stars, and thus should be included for a proper description of most of the evolutionary phases of low-mass stars. In PGPUC, we used the opacity tables calculated by \citet[][F05]{Ferguson_etal2005}, instead of the opacity tables calculated by Bell (1994, unpublished) for the PG SEC using the SSG synthetic spectrum program \citep{Bell_Gustafsson1978, Gustafsson_Bell1979}. The F05 tables were obtained with a modified version of the stellar atmosphere code PHOENIX \citep{Alexander_Ferguson1994}, which includes an improved equation of state, additional atomic and molecular lines, and dust opacity.

\subsection{Boundary Conditions}

When a stellar model is computed, boundary conditions are required to solve the problem. In the center of a star the conditions are that mass and luminosity are zero, while in the outer layers a relation between the temperature and the optical depth ($\tau$) is required to know the behavior of the temperature, and then to obtain the pressure. 

In a simplified treatment, this relation can be obtained analytically, as \citet{Eddington1926} did, where $ T^4(\tau) = \frac{3}{4} T^4_{\rm eff} \left( \tau + \frac{2}{3} \right)$, which defines the effective temperature ($T_{\rm eff}$) as the temperature where a photon emitted from the depth with $\tau = 2/3$ has a probability of $e^{-2/3} \sim 0.5$ of emerging from the surface. However, this relation uses the gray atmosphere approximation, that is not a bad approximation when: i)~the opacity is dominated by H$^-$, ii)~local thermodynamical equilibrium is achieved (which is not appropriate for low-surface-gravity stellar atmospheres, where densities are too low), iii)~hydrostatic equilibrium is valid (the pressure gradient exactly balances gravity, which is not valid for hot stars), and iv)~there is radiative equilibrium (radiation is the main form of energy transport). 

On the other hand, reliable photospheric boundary conditions can be determined without these approximations from actual stellar atmosphere models \citep[e.g.,][]{VandenBerg_etal2008, Dotter_etal2008}, but these calculations need a lot of computing time, and due to the many parameters involved, tabulated tables covering the full evolutionary track of a GC star do not yet exist. Instead of using a theoretical relation with many approximations, or one that is very time consuming, semi-analytical formulae are often used. PG SEC uses the \citet{KrishnaSwamy1966} T-$\tau$ relation, which was determined using the limb darkening of the Sun. However, \citet{Catelan2007} has pointed out that this relation does not fit correctly the Krishna Swamy data, suggesting the following improved fit \footnote{Catelan's published formula was misprinted.}:

\begin{equation}
\label{Cat07}
T^4(\tau) = \frac{3}{4} T^4_{\rm eff} \left( 0.55 + 5.46\tau - 7.72 \tau^{1.5} + 5.12 \tau^{2} - 1.08 \tau^{2.5}  \right),
\end{equation}

\noindent which is now used in the PGPUC SEC.

\subsection{Thermonuclear reaction rates}

The thermonuclear reaction rates taken into account in the PG SEC are the reactions related to hydrogen and helium burning, as given by \citet{Caughlan_Fowler1988} except for the higher rate of the $^{12}$C($\alpha$,$\gamma$)$^{16}$O reaction advocated by \citet{Weaver_Woosley1993}. In this update, all the reaction rates were changed from the analytical formulas determined by \citet{Caughlan_Fowler1988} to the latest available analytical formulae of the NACRE compilation \citep{Angulo_etal1999}, where each rate is determined with an extrapolation of the measured cross section down to astrophysical energies, following theoretical approximations. However, for two important reactions, namely $^{12}$C($\alpha$,$\gamma$)$^{16}$O and $^{14}$N(p,$\gamma$)$^{15}$O, the rates used are from \citet[][K02]{Kunz_etal2002} and \citet[][F04]{Formicola_etal2004}\footnote{The analytical formula was determined by \citet{Imbriani_etal2005}.}, respectively.

\subsection{Equation of state}

To determine the structure and evolution of a star, it is required to solve the four basic equations of structure. However, an additional equation is needed to solve this set of equations, namely the so-called equation of state (EOS), which determines the density ($\rho$) knowing the pressure ($P$), the temperature ($T$), and the chemical composition. In this work we changed the original EOS \citep[][]{Sweigart1973} to Irwin's EOS \citep[FreeEOS,][]{Irwin2007}\footnote{All FreeEOS documentation is available at \url{http://freeeos.sourceforge.net/}.}, which takes into account non-ideal effects, mainly pressure ionization, Coulomb effects, and quantum electrodynamic corrections, which can be important even at non-relativistic temperatures and non-degenerate densities \citep{Heckler1994}. The 20 elements considered in FreeEOS are H, He, C, N, O, Ne, Na, Mg, Al, Si, P, S, Cl, A, Ca, Ti, Cr, Mn, Fe, and Ni, while all molecules are ignored except for H$_2$ and H$_2^+$. Because of the several time-consuming non-ideal effects included in FreeEOS, different approximations have been implemented to solve the equations, reducing the CPU time required. The most accurate FreeEOS option (EOS1) takes into account 295 ionization states of the 20 elements, while EOS1a considers some less abundant metals as fully ionized. EOS2, EOS3, and EOS4 use a faster (but less accurate) approximation for the pressure ionization effect; however, while EOS2 considers the 295 ionization states, EOS3 and EOS4 consider all metals fully ionized when $\log T >$ 6. The difference between EOS3 and EOS4 is that the latter uses the approximation of \citet[][P95]{Pols_etal1995} to calculate the number of particles used to determine the Coulomb free energy, and it is exact when all elements are neutral or completely ionized. However, the P95 algorithm has a discontinuity problem when H$_2$ is dissociated, which affects mainly stars with masses below 0.1 $M_\odot$. In PGPUC we use the EOS4 approximation due to its lowest consuming time, and the small difference in almost all the parameters with the other approximations (see Table \ref{TableFreeEOS2}). As can be seen, the mass of the helium core ($M_{cHe}$), one of the most important parameters that defines the whole HB phase, only changes by 0.002\% with respect to the calculation with EOS1.

\begin{table*}
\center
  \begin{tabular}{|cccccccccc|}
\hline
EOS& CPU   & $L_\nu$ & $L$         & $T_{\rm eff}$& $R$     & Age       & $M_{cHe}$ & $T_c$                 & $\rho_c$\\
   & (days)& ($L_\odot$)  & ($L_\odot$) & (K)          & (10$^{12}$cm)    & (Myr)     & ($M_\odot$) & (10$^{7}$K)    & (10$^5$ g/cm$^3$)\\
\hline
1  & 12& 1.00342  & 2013.1 & 3925.7 & 6.758 & 13162.68 & 0.47977 & 7.4432 & 9.931\\
3  & 0.336   & 1.00384  & 2015.3 & 3925.2 & 6.763 & 13162.73 & 0.47978 & 7.4435 & 9.933\\
4  & 0.156   & 1.00348  & 2014.8 & 3925.4 & 6.762 & 13162.75 & 0.47976 & 7.4442 & 9.932\\
\hline
   \end{tabular}
	\caption{Properties of a 0.8 $M_\odot$ star (Y=0.25, Z=0.001, [$\alpha$/Fe]=0.0) at the RGB tip, where the complete evolution was computed using the labeled FreeEOS approximation (in case of EOS1a and EOS2, PGPUC SEC did not converge with these initial properties). The other columns are: CPU time used to calculate the whole evolution from the MS to the RGB tip (based on three Intel Xeon E5405 2.0 GHz processors, in parallel-processing mode), luminosity via neutrino emission ($L_\nu$), total luminosity ($L$), effective temperature ($T_{\rm eff}$), stellar radius ($R$), the time spent from the MS to the RGB tip (Age), the mass of the helium core ($M_{cHe}$), central temperature ($T_c$), and central density ($\rho_c)$.}
	\label{TableFreeEOS2}
\end{table*}

\subsection{Mass Loss}

Historically, the mass loss rate has usually been determined by the formula of \citet[][R75]{Reimers1975}, which was obtained empirically. It reads 

\begin{equation}
\dot{M} = 4\times 10^{-13} \eta_R \frac{L_* R_*}{M_*} M_\odot/yr,
\label{Reimers}
\end{equation}

\noindent where $L_*$, $R_*$, and $M_*$ are the luminosity, radius, and mass of the star in solar units, respectively, and $\eta_R$ is a free parameter whose value is usually taken as $\sim$0.5 due to uncertainty in the numerical coefficient (e.g., in order to fit the observed HB morphology in the GC's the numerical coefficient used is $2\times 10^{-13}$). It is now well known that this formula does not provide a good description of the available mass loss rates for RGB stars, and several alternative prescriptions have been proposed in the literature \citep[e.g.,][ and references therein]{Catelan2000,Catelan2009}. In particular, \citet[SC05]{Schroder_Cuntz2005} have recently suggested a new mass loss rate formula that more properly describes the empirical data and additional physical constraints. They obtained

\begin{equation}
\label{SC05}
\dot{M} = 8\times 10^{-14} \eta \frac{L_* R_*}{M_*} \left(\frac{T_{\rm eff}}{4000 K}\right)^{3.5}\left( 1 + \frac{g_\odot}{4300 g_*}\right) M_\odot/yr,
\end{equation}

\noindent where $L_*$, $R_*$, and $M_*$ have the same meaning as in eq. (\ref{Reimers}), $T_{\rm eff}$ is the effective temperature in Kelvin, $g_\odot$ is the solar surface gravity, $g_*$ is the stellar surface gravity, and $\eta$ is a free parameter that we added to allow varying the final stellar mass, even though the value $8\times 10^{-14}$ was calibrated to satisfy the HB morphology of two GCs (NGC 5904=M5 and NGC 5927), suggesting that the $\eta$ value added here is typically $\sim$1 (SC05). The two new parameters ($T_{\rm eff}$ and $g_*$) are related to the mechanical energy flux added to the outer layer of the star (depends on $T_{\rm eff}$) and to the chromospheric height (depends on $g_*$). This new formula has been tested using well-studied nearby stars with measured mass loss rates, reproducing them within the error bars \citep[][]{Schroder_Cuntz2007}. For these reasons, we used the SC05 mass loss rate formulation, instead of the R75 one.

\begin{figure}
  \centering
     \includegraphics[height=9cm, trim=0.7cm 1cm 0cm 0.5cm]{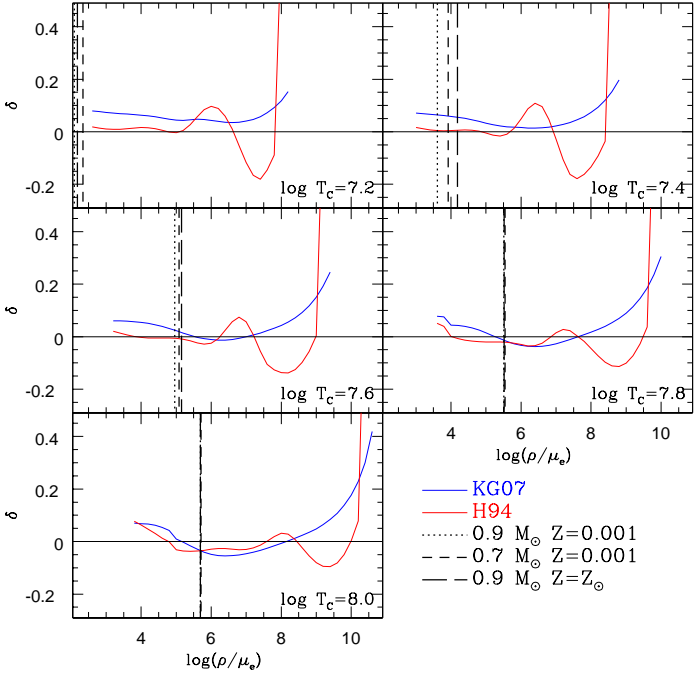}
     \caption{Fractional error $\delta =(Q_{\rm fit}-Q_{\rm numKG07})/Q_{\rm numKG07}$ for temperatures concerning low-mass RGB stars, where $Q_{\rm fit}$ is the neutrino emissivity calculated with the H94 (red lines) or KG07 (blue lines) fitting formulae, and $Q_{\rm numKG07}$ is the neutrino emissivity calculated numerically by KG07. Vertical black lines indicate the $\log(\rho/\mu_e)$ value in the center of three stars (Y=0.245, and the labeled initial mass and metallicity) along their evolution through the RGB phase, when the center reaches the labeled temperature. In the panel for $\log T = 8$, vertical lines indicate the density for the maximum temperature reached in the center ($\log T \sim 7.87$). The $\log(\rho/\mu_e$) ranges are determined by the limits of the KG07 numerical values.}
     \label{CompNeutrino}
\end{figure}

\subsection{Neutrino energy loss}

When a photon is propagating in a plasma under sufficiently extreme conditions (high temperature, high density), the photon can be coupled to this plasma. When this happens, the photon can decay into a neutrino and an anti-neutrino (emitted in opposite directions), which cannot occur when a photon is not coupled to the plasma, since energy and momentum cannot simultaneously be conserved under these circumstances \citep[e.g.,][KG07]{Kantor_Gusakov2007}. In low-mass stars, these extreme conditions are reached in the degenerate core when the star is in the RGB and the white dwarf phases. Generally, in stellar evolution codes fitting formulae are used to take into account the energy loss by this effect, the most common ones being those proposed by \citet[][1995, H94]{Haft_Raffelt_Weiss1994} and \citet[][I96]{Itoh_etal1996}. However, KG07 have pointed out the ineffectiveness of these formulae to reproduce the real energy loss via neutrino emission for extreme values of $\rho/\mu_e$ while, moreover, the I96 fitting formula is only valid near the maximum of emissivity. KG07 obtained this emissivity numerically, suggesting a new fitting formula, which solves the problem for extreme values.

Figure \ref{CompNeutrino} shows the fractional error of the H94 and KG07 fits, compared to the numerical neutrino emissivity from KG07. It can be seen that the H94 formula reproduces the numerical results better in the relevant conditions for low-mass stars (vertical black lines). For this reason, we continue using the H94 recipe. However, if calculations for extreme values are required, we recommend the use of an interpolation of their numerical tables.

\subsection{Comparison with other codes}

In this subsection we compare the PGPUC SEC with some SECs widely used in the study of low-mass stars and GCs. These SECs are: Victoria-Regina \citep[][hereafter VIC-REG]{VandenBerg_etal2006}, BaSTI \citep[][]{Pietrinferni_etal2004}, GARSTEC \citep{Weiss_Schlattl2008}, STAREVOL \citep{Siess2006}, and PADOVA \citep{Bertelli_etal2008}. Table \ref{TabComp} shows the input physics used in these codes, where it can be seen that BaSTI and GARSTEC include nearly the same physics used in PGPUC.

\begin{figure}[!t]
   \centering
      \includegraphics[height=7.7cm, trim=0cm 3.3cm 0cm 0cm]{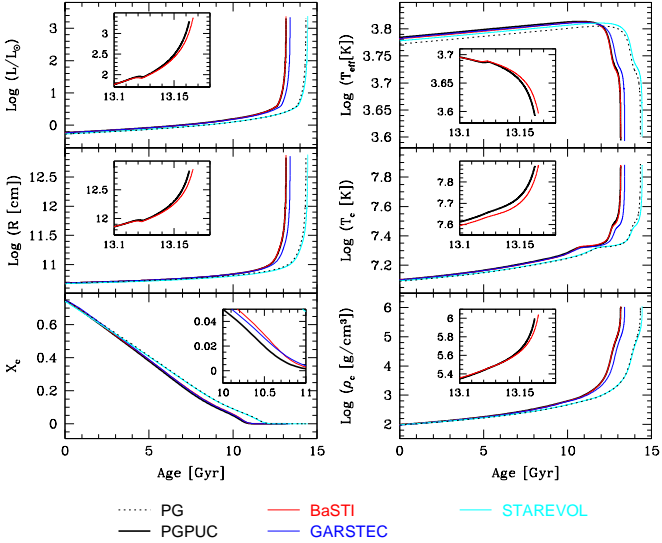}
      \caption{Comparison of some evolutionary properties for a 0.8 $M_\odot$ star with Y=0.250 and Z=0.001, as predicted by some of the codes presented in Table \ref{TabComp}: PG, dotted black lines; PGPUC, black lines; BaSTI, red lines; GARSTEC, blue lines; STAREVOL, cyan lines. The properties are luminosity ($L/L_\odot$), effective temperature ($T_{\rm eff}$), radius ($R$), central temperature ($T_c$), density ($\rho_c$), and hydrogen abundance ($X_C$). Small boxes are zoom-in to show the small differences between BaSTI and PGPUC results.}
      \label{CompAll_age_y25}
 \end{figure}

In Figure \ref{CompAll_age_y25}, we compare some stellar (internal and external) properties of the SECs that have these data available\footnote{GARSTEC evolutionary tracks were obtained from the Stellar Model Challenge project (\href{http://www.mpa-garching.mpg.de/stars/SCC/Models.html}{http://www.mpa-garching.mpg.de/stars/} \href{http://www.mpa-garching.mpg.de/stars/SCC/Models.html}{SCC/Models.html}), while PADOVA and VIC-REG ones were download from their respective web pages: \href{http://www3.cadc-ccda.hia-iha.nrc-cnrc.gc.ca/community/VictoriaReginaModels}{http://www3.cadc-ccda.hia-iha.} \href{http://www3.cadc-ccda.hia-iha.nrc-cnrc.gc.ca/community/VictoriaReginaModels}{nrc-cnrc.gc.ca/community/VictoriaReginaModels} and \href{http://pleiadi.pd.astro.it/}{http://pleiadi.} \href{http://pleiadi.pd.astro.it/}{pd.astro.it}. STAREVOL evolutionary track was obtained via private communication with Ana Palacios. BaSTI evolutionary tracks were obtained from the official webpage \href{http://albione.oa-}{http://albiones.oa-teramo.inaf.it/} or via private communication with Santi Cassisi.}. The star has a mass of $0.8 M_\odot$, an initial helium abundance of ${\rm Y} = 0.250$ \citep[near to the primordial helium abundance ${\rm Y_p} = 0.245$;][and references therein]{Izotov_etal2007}, a metal abundance of Z=0.001, a solar scaled abundance of alpha elements (${\rm [\alpha /Fe] = 0.0}$) according to \citet{Grevesse_Noels1993}, and the evolution is without mass loss. These properties are close to the mass, helium, and metallicity of turn-off (TO) stars in GCs. Moreover, only for this comparison PGPUC uses the NACRE reaction rate for $^{14}$N(p,$\gamma$)$^{15}$O, which is the rate used for all codes that use NACRE as input. In that figure the good agreement of PGPUC with BaSTI and GARSTEC can be observed, the latter being the codes using FreeEOS (interpolated tables). Note that the remaining different physical inputs could possibly produce the observed differences; e.g., if the GARSTEC SEC adopted the conductive opacities from C07 instead of \citet{Itoh_etal1984}, its predicted luminosity at the end of the RGB phase would decrease due to the earlier helium flash (C07). The differences with the PG and STAREVOL SECs must be produced by the different EOS's which lead to a lower central density ($\rho_c$) and temperature ($T_c$) than the other three codes, and, consequently, a lower hydrogen burning rate.

\begin{table*}
\center

  \begin{tabular}{| c | c c c c c c c |}
\hline
    		& Princeton-	& Princeton-	& Victoria-	& BaSTI    	& GARSTEC & STAREVOL & PADOVA	\\
    Physics	& Goddard	& Goddard-PUC	& Regina	&		&	  &	     & 		\\ 
		& (PG)		& (PGPUC)	&		&		&	  &	     & 		\\ 
\hline
 Opacity Tables	&		&		&		&		&	  &	     &		\\ 
High Temp	& OPAL92	& OPAL96	& OPAL92	& OPAL96	& OPAL96  & OPAL96   & OPAL96	\\ 
Low Temp	& B94		& F05		& AF94		& AF94		& F05	  & AF94     & AF94	\\ 
Conductive	& HL69‏+C70 	& C07		& HL69		& P99       	& I83  	  & HL69     & I83	\\ 
\hline
 Neutrino	&		& 		&		&		&	  &	     &		\\ 
Energy Loss	& H94		& H94		& I96		& H94		& H94	  & I96	     & H94	\\ 
\hline
 Reaction Rates	& 		& 		&		& 		&	  & 	     &		\\ 
$^{12}$C($\alpha$,
$\gamma$)$^{16}$O & WW93	& K02		& BP92		& K02		& K02     & NACRE    & WW93	\\ 
Other		& CF88		& NACRE		& BP92		& NACRE		& NACRE	  & NACRE    & CF88	\\ 
\hline
Equation of 	& 		& 		&		& 		&	  &	     &		\\
State		& S73		& FreeEOS	& VB00		& FreeEOS	& FreeEOS & P95	     & M90+G96	\\ 
\hline
Boundary	& 		& 		&		& 		&	  &	     &		\\ 
Condition	& KS66		& Cat07 (our eq. \ref{Cat07})& KS66	& KS66		& KS66	  & KS66     & KS66	\\ 
\hline
Mass Loss 	& R75		& SC05		& R75		& R75		& R75	  & R75      & R75	\\ 
\hline
$\alpha_{MLT}$ (GN93)& 1.845 	& 1.886		& 1.890		& 1.913		& 1.739	  & 1.853    & 1.68(GN98)\\
\hline
  \end{tabular}
\caption{Physical inputs included in compared SECs. \textbf{References}: AF94 \citep{Alexander_Ferguson1994}, BP92 \citep{Bahcall_Pinsonneault1992}, B94 (Bell 1994, unpublished), C70 \citep{Canuto1970}, C07 \citep{Cassisi_etal2007}, Cat07 \citep{Catelan2007}, CF88 \citep{Caughlan_Fowler1988}, FreeEOS \citep{Irwin2007}, F05 \citep{Ferguson_etal2005}, G96 \citep{Girardi_etal1996}, HL69 \citep{Hubbard_Lampe1969}, H94 \citep{Haft_Raffelt_Weiss1994}, I83 \citep{Itoh_etal1984}, I96 \citep{Itoh_etal1996}, K02 \citep{Kunz_etal2002}, KS66 \citep{KrishnaSwamy1966}, M90 \citep{Mihalas_etal1990}, NACRE \citep{Angulo_etal1999}, OPAL92 \citep{Rogers_Iglesias1992}, OPAL96 \citep{Iglesias_Rogers1996}, P95 \citep{Pols_etal1995}, P99 \citep{Potekhin1999}, R75 \citep{Reimers1975}, S73 \citep{Sweigart1973}, SC05 \citep{Schroder_Cuntz2005}, VB00 \citep{Vandenberg_etal2000}, WW93 \citep{Weaver_Woosley1993}. \textbf{Chemical compositions}: GN93 \citep[$Z/X=0.0244$,][]{Grevesse_Noels1993}, GS98 \citep[$Z/X=0.0231$,][]{Grevesse_Sauval1998}.}

\label{TabComp}
\end{table*}

\begin{figure*}
\begin{minipage}[l]{0.333333\linewidth}
\centering
\includegraphics[height=6.1cm, trim=1.4cm 0cm 1cm 1cm]{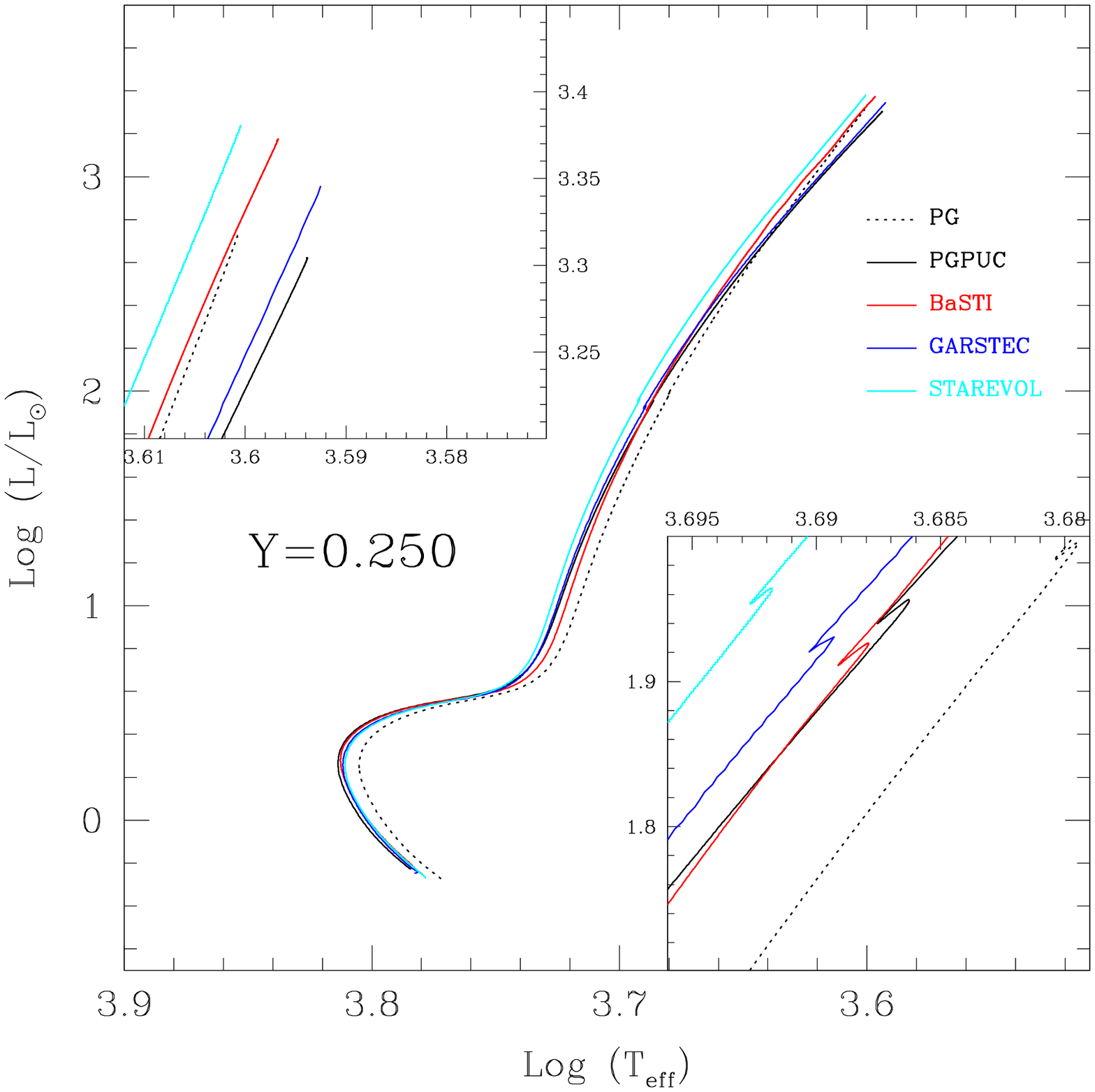}
\end{minipage}
\hspace{0.1cm}
\begin{minipage}[r]{0.333333\linewidth}
\centering
\includegraphics[height=6.1cm, trim=0.7cm 0cm 1cm 1cm]{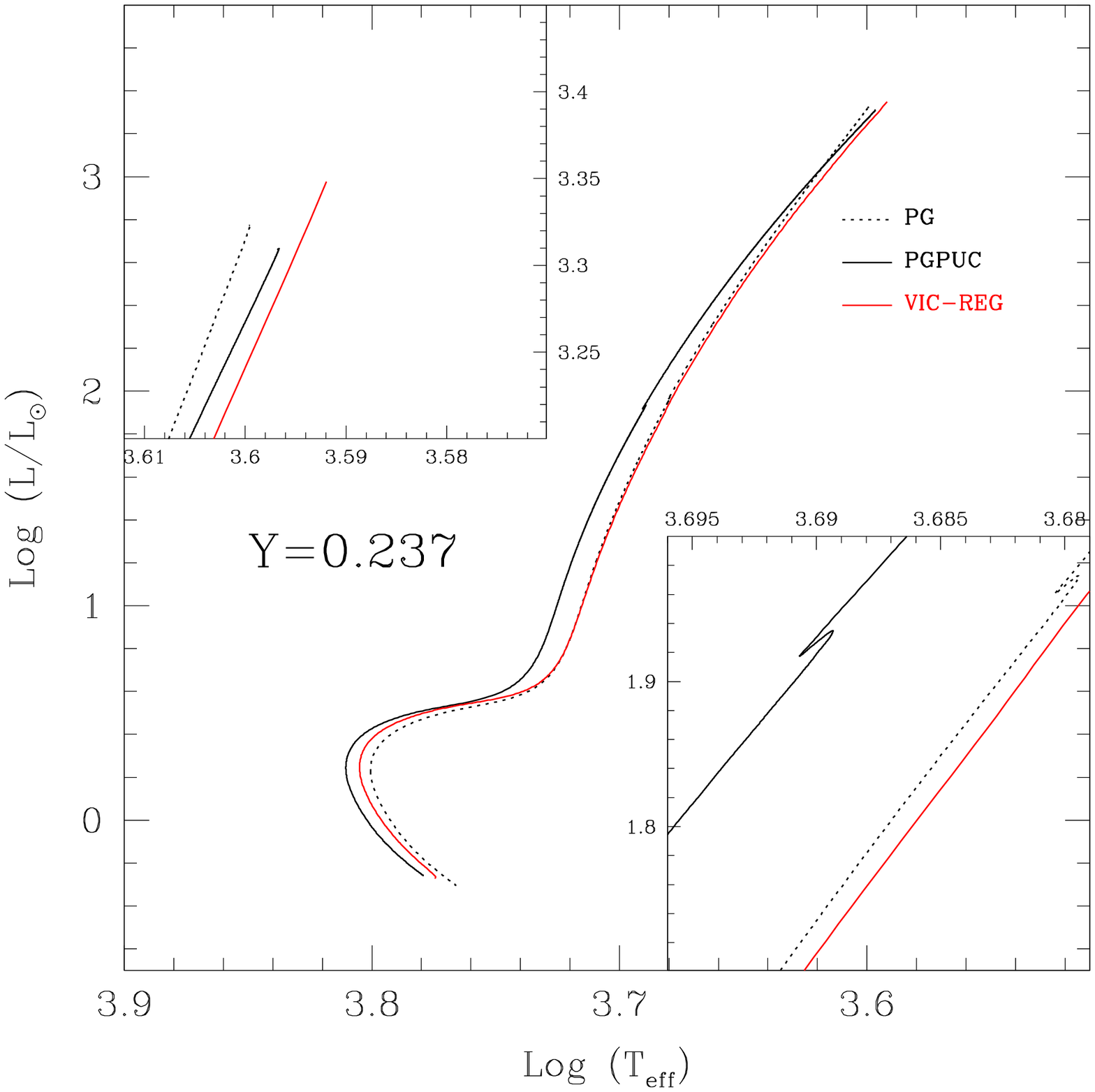}
\end{minipage}
\hspace{0.1cm}
\begin{minipage}[r]{0.333333\linewidth}
\centering
\includegraphics[height=6.1cm, trim=1cm 0cm 1cm 1cm]{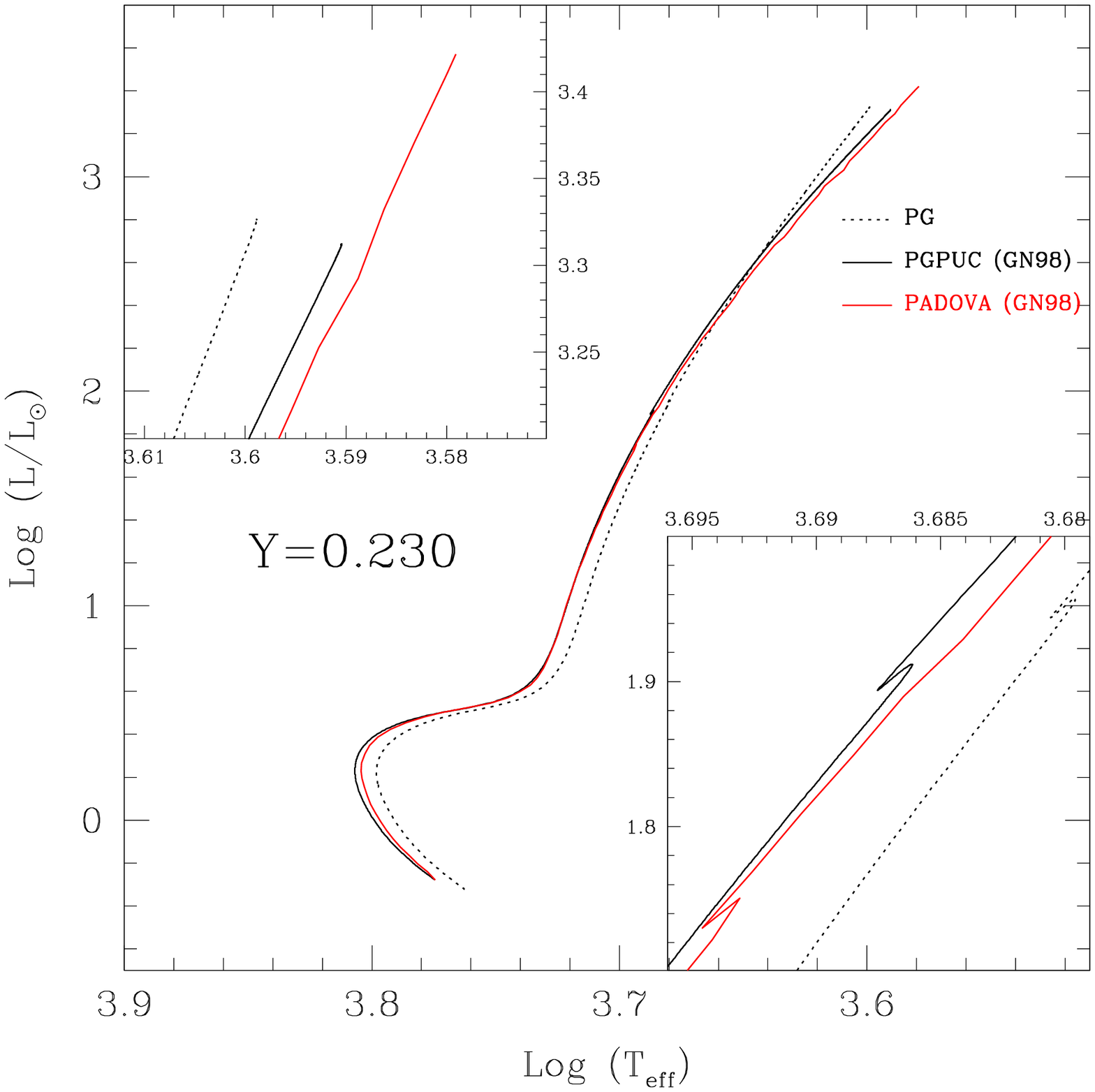}
\end{minipage}
\caption{RGB evolutionary track comparison between PGPUC and the labeled codes (see Table \ref{TabComp}) for a 0.8 $M_\odot$ star with three different helium abundances: Y=0.250 (left), Y=0.237 (middle) and Y=0.230 (right). The metallicity is Z=0.001 and the chemical composition is solar-scaled according to \citet{Grevesse_Noels1993}, except when labeled GN98, where the chemical composition is solar-scaled according to \citet{Grevesse_Sauval1998}. Each panel contains an upper left and a lower right panel with zooms to the RGB tip and RGB bump regions, respectively, with the same scales for each Y.}
\label{FigComparison}
\end{figure*}

\begin{figure*}
\begin{minipage}[l]{0.333333\linewidth}
\centering
\includegraphics[height=6.1cm, trim=1.4cm 0cm 1cm 1cm]{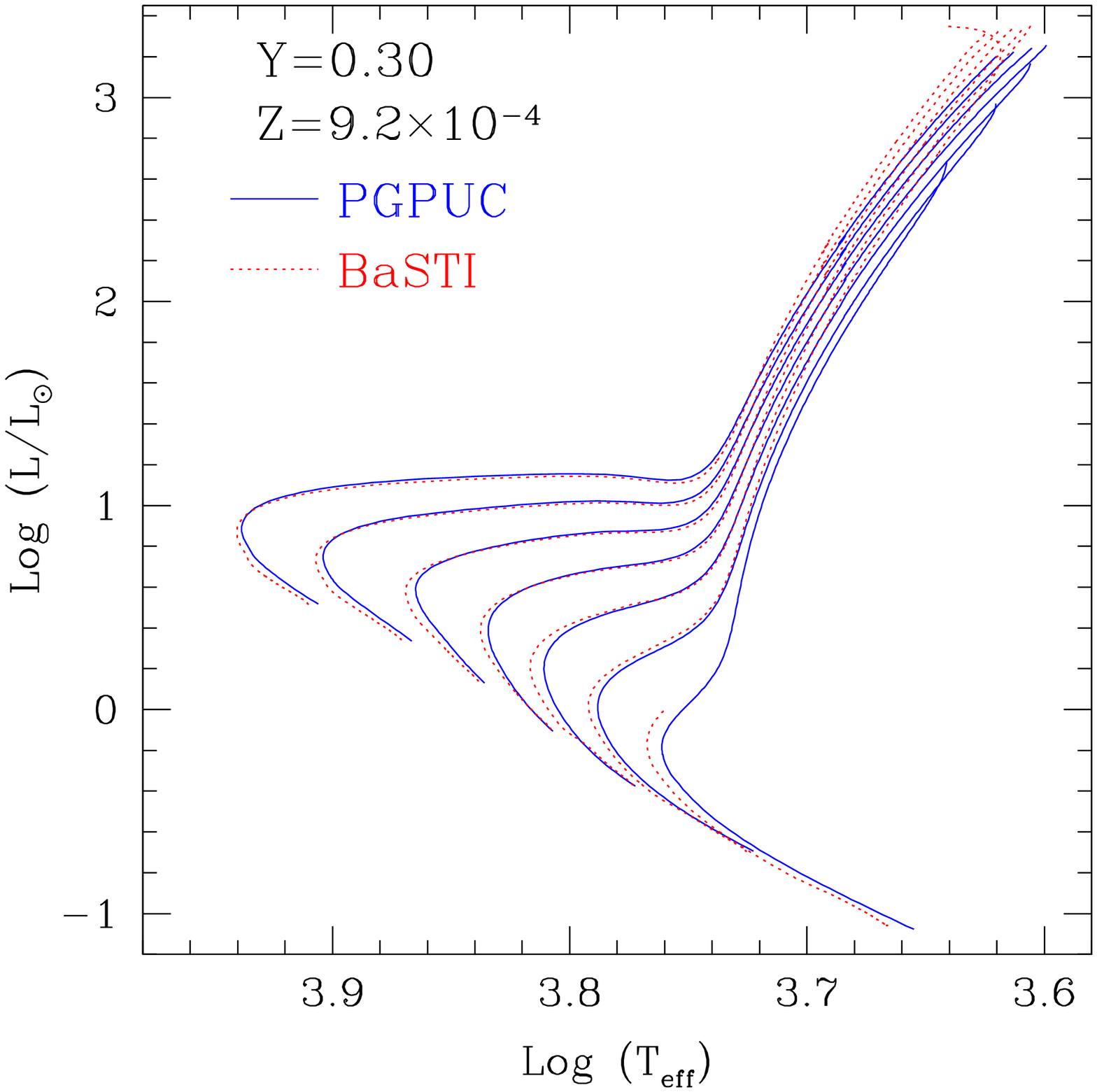}
\end{minipage}
\hspace{0.1cm}
\begin{minipage}[r]{0.333333\linewidth}
\centering
\includegraphics[height=6.1cm, trim=0.7cm 0cm 1cm 1cm]{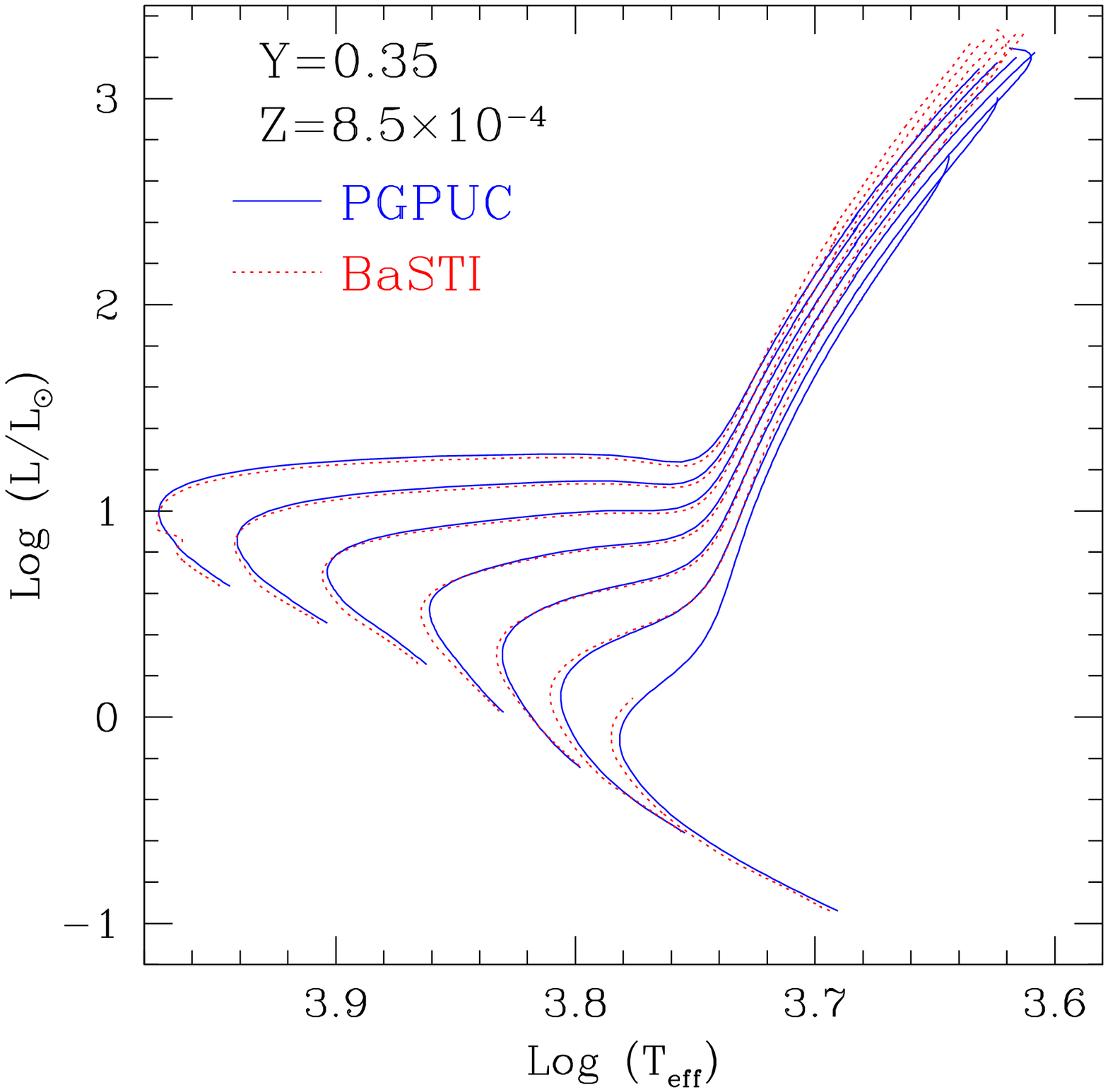}
\end{minipage}
\hspace{0.1cm}
\begin{minipage}[r]{0.333333\linewidth}
\centering
\includegraphics[height=6.1cm, trim=1cm 0cm 1cm 1cm]{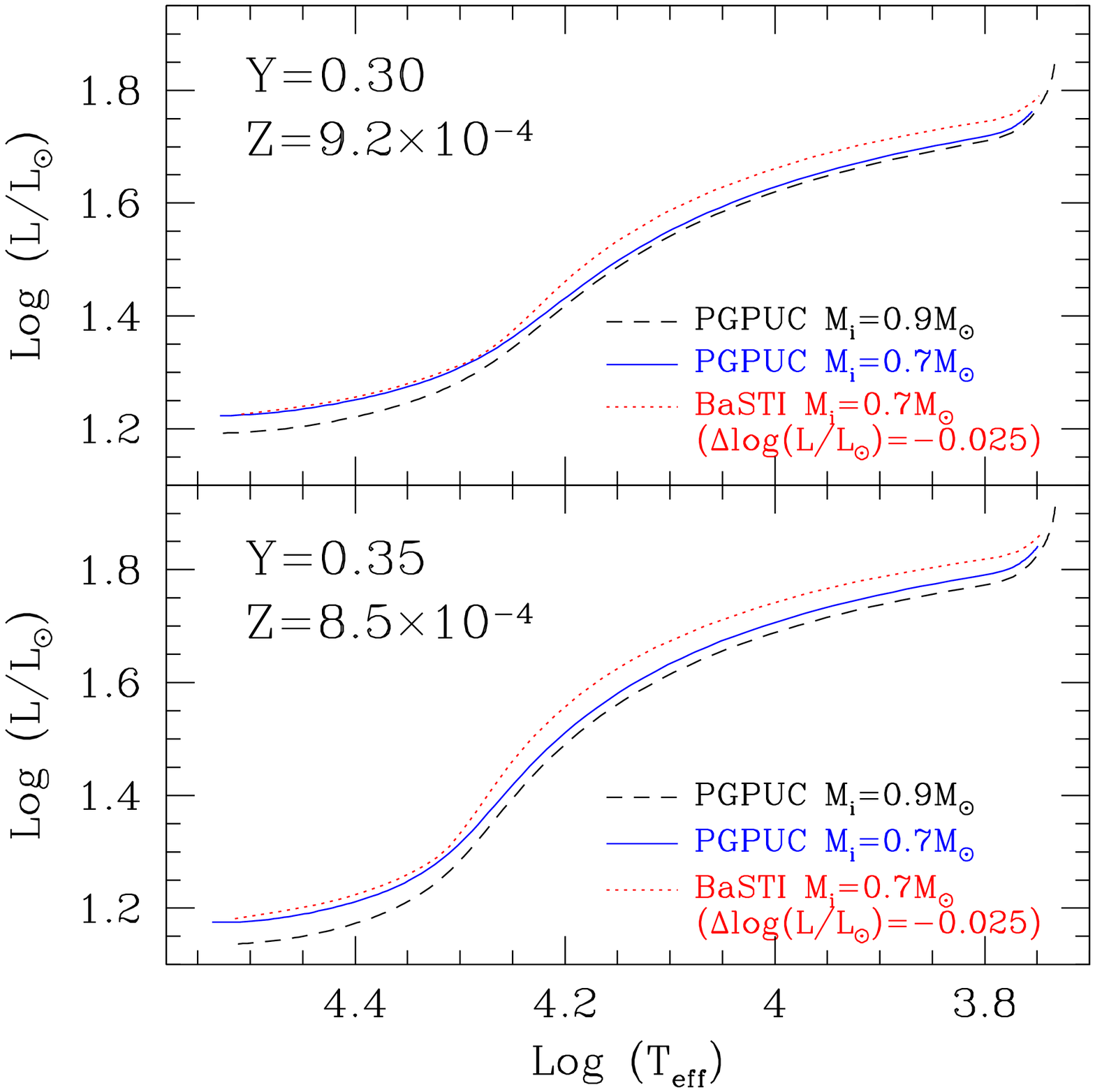}
\end{minipage}
\caption{Comparison between PGPUC and BaSTI results for non-standard chemical compositions. Left panel: Evolutionary tracks with ${\rm Y}=0.30$ and ${\rm Z}=9.2\times10^{-4}$. Middle panel: Evolutionary tracks with ${\rm Y}=0.35$ and ${\rm Z}=8.5\times10^{-4}$. Right panels: ZAHB loci for ${\rm Y}=0.30$ and ${\rm Z}=9.2\times10^{-4}$ (upper panel) and  ${\rm Y}=0.35$ and ${\rm Z}=8.5\times10^{-4}$ (lower panel), where for comparison BaSTI ZAHB loci are moved $\Delta\log(L/L_\odot)=-0.025$. PGPUC ZAHB loci are shown for progenitor masses of $M_i=0.7M_\odot$ and $M_i=0.9M_\odot$.}
\label{FigCompBaSTI}
\end{figure*}

The impact of these differences can also be observed in an HR diagram, 
where in the left panel of Fig. \ref{FigComparison} we show the evolutionary tracks calculated with the same codes as in Fig. \ref{CompAll_age_y25}. As can be observed, PGPUC has better agreement in the MS with BaSTI, GARSTEC, and STAREVOL, but PGPUC produces the hottest MS ($\Delta T_{\rm eff} \sim 35$ K at a solar luminosity). In the sub-giant branch (SGB) these differences are reduced, but at the base of the RGB STAREVOL begins to diverge from the other three codes. At the RGB bump PGPUC is slightly brighter and cooler than predicted by BaSTI and GARSTEC ($\Delta T_{\rm eff} \sim 18$ K and $\Delta \log (L/L_\odot)\sim 0.03$ with BaSTI). Finally, at the RGB tip the PGPUC track has the lower luminosity. 
Although there are several other differences along the RGB evolution, they tend to be very small.

For comparison with the VIC-REG model (middle panel of Fig. \ref{FigComparison}), we produced an additional evolutionary track with the same properties as used to produce the left panel of the same figure, except for the initial helium abundance that in this case is ${\rm Y = 0.237}$. However, due to the different input physics between PGPUC and VIC-REG, the differences along the evolution are relatively large\footnote{During the publication of this paper, an updated version of VIC-REG was published by \citet{VandenBerg_etal2012}, which could reduce these differences.}. In the right panel of Fig. \ref{FigComparison}, we also compared with PADOVA evolutionary track, where the properties are: ${\rm M = 0.8} M_\odot$, ${\rm Y = 0.230}$, ${\rm Z = 0.001}$, and ${\rm [\alpha /Fe] = 0.0}$ according to \citet{Grevesse_Sauval1998}. It reveals some small differences similar to those in the left panel, but at the RGB bump there is a difference of $\Delta \log (L/L_\odot)\sim$0.15 (PGPUC is brighter). Also, at the RGB tip the difference has the same magnitude, but PGPUC is instead fainter. This could be due to the differences in the conductive opacity tables and in the thermonuclear reaction rates, because both changes produced an increase of the RGB bump luminosity and a decrease of the RGB tip luminosity. 

Due to the fact that this paper is focused on He-enhanced theoretical models, in Fig. \ref{FigCompBaSTI} we compared PGPUC and BaSTI results for ${\rm [Fe/H]}=-1.62$ with Y=0.30 and Y=0.35, where PGPUC results are interpolated as is described in Appendix \ref{AppPGPUConline} \citep[see also][]{Dotter_etal2008, Bertelli_etal2009, DiCriscienzo_etal2010, DellOmodarme_etal2012}. In the left and middle panels, we show the comparison of evolutionary tracks with masses between 0.5 and 1.1$M_\odot$ in intervals of $\Delta M=0.1 M_\odot$. As can be seen, there are not large differences at the MS except for 0.7 and 0.5$M_\odot$, where PGPUC tracks are cooler. At high luminosities it is possible to observe that PGPUC results are also cooler and fainter than BaSTI tracks. In the right panels are shown the comparisons between the ZAHB loci of the two SECs, which are created using a progenitor mass of 0.7$M_\odot$. For this comparison we have to move the BaSTI ZAHB loci by $\Delta\log(L/L_\odot)=-0.025$, since the BaSTI's code uses the P99 conductive opacity table inducing a brighter ZAHB than the C07 ones (see C07). Apart from previous difference in luminosity, there is still a difference at the red part of the ZAHB, with the most important factor likely being the rate of the reaction $^{14}$N(p,$\gamma$)$^{15}$O that is adopted in both codes. More specifically, for this comparison the PGPUC SEC uses the F04 rate, whereas BaSTI uses the NACRE one. \citet{Pietrinferni_etal2010} pointed out that two ZAHB loci created with these rates have a difference in luminosity of the order presented here at $\log T_{\rm eff}=3.85$, the ZAHB determined with the NACRE reaction rate being brighter (at this metallicity). As far as the predicted temperatures are concerned, we do not find any significant differences. We also show the ZAHB loci created with a progenitor mass of 0.9$M_\odot$, and this case is discussed in more detail in section \ref{ZAHBage}. Finally, we conclude that PGPUC models are in agreement with BaSTI ones, except for ZAHB loci being fainter by $\Delta\log(L/L_\odot)=-0.025$, due to the differences in the adopted input physics.

\section{PGPUC database}
\label{PGPUCdatabase}

\begin{table*}[t]
\center
\begin{tabular}{c|ccccccc|}
\cline{2-8}
       &       \multicolumn{7}{c|}{Y}  \\ 
\hline
\multicolumn{1}{|c|}{Z}       &  0.230 & 0.245       & 0.270 & 0.295 & 0.320 & 0.345 & 0.370 \\
\hline
\multicolumn{1}{|c|}{1.60e-4} & -2.259 & {\bf  -2.250} & -2.235 & -2.220 & -2.205 & -2.188 & -2.171\\
\multicolumn{1}{|c|}{2.80e-4} & -2.015 & {\bf  -2.007} & -1.992 & -1.977 & -1.961 & -1.945 & -1.928\\
\multicolumn{1}{|c|}{5.10e-4} & -1.755 & {\bf  -1.746} & -1.732 & -1.717 & -1.701 & -1.685 & -1.668\\
\multicolumn{1}{|c|}{9.30e-4} & -1.494 & {\bf  -1.485} & -1.471 & -1.455 & -1.440 & -1.423 & -1.407\\
\multicolumn{1}{|c|}{1.60e-3} & -1.258 & {\bf  -1.249} & -1.235 & -1.219 & -1.204 & -1.187 & -1.170\\
\multicolumn{1}{|c|}{2.84e-3} & -1.008 & {\bf  -0.999} & -0.985 & -0.969 & -0.954 & -0.937 & -0.920\\
\multicolumn{1}{|c|}{5.03e-3} & -0.758 & {\bf  -0.750} & -0.735 & -0.720 & -0.704 & -0.688 & -0.671\\
\multicolumn{1}{|c|}{8.90e-3} & -0.508 & {\bf  -0.500} & -0.485 & -0.470 & -0.454 & -0.437 & -0.420\\
\multicolumn{1}{|c|}{1.57e-2} & -0.258 & {\bf  -0.249} & -0.234 & -0.219 & -0.203 & -0.186 & -0.169\\
\hline
   \end{tabular}
	\caption{Iron abundances [Fe/H] as a function of the metallicity value Z and the helium abundance Y for an $\alpha$-element enhancement of [${\rm \alpha }$/Fe]=0.3. In boldface type are the selected [Fe/H]'s (when Y = 0.245) to obtain the Z values used to calculate our theoretical models.}
	\label{TableChemical}
\end{table*}

\subsection{Evolutionary Tracks}

To study the effects of a helium enrichment in globular clusters, we created a database of evolutionary tracks covering the evolutionary phases from the zero-age main sequence (ZAMS) to the tip of the RGB, together with zero-age horizontal branch (ZAHB) loci. This database of evolutionary tracks includes stellar masses between 0.5 and 1.1 $M_\odot$ with $\Delta M = 0.1$ $M_\odot$, for 7 helium abundances, 9 metallicities, an enhancement of the alpha elements by $[\alpha/{\rm Fe}] = 0.3$, and a mass loss rate according to equation \ref{SC05} (with $\eta = 1.0$). The metallicity values were selected to obtain the iron abundances [Fe/H] shown in the second column of Table \ref{TableChemical} when ${\rm Y}=0.245$. Since it is assumed that nearly all GCs do not undergo internal chemical enrichment due to supernova explosions, the mass fraction of Fe within a given GC (but not [Fe/H]) would be constant over the range of possible helium abundances \citep[e.g.,][]{Valcarce_Catelan2011}. For this reason we will use the same metallicity value for each set of Y values in our analysis. Even though variations of other elements can be associated to He-enrichment and also produce some changes in the theoretical evolutionary tracks and ZAHB loci \citep[C+N+O and/or other heavier elements;][]{Milone_etal2008, Cassisi_etal2008, Pietrinferni_etal2009, Ventura_etal2009, MarinFranch_etal2009, AlvesBrito_etal2012, VandenBerg_etal2012}, in this paper we focus exclusively on He. We leave the calculations with non-standard metal ratios for future papers, where we also plan to increase the ranges of masses, metallicities, and $[\alpha/{\rm Fe}]$ ratios covered.

We adopted the solar chemical abundances presented by \citet[][GS98]{Grevesse_Sauval1998}, where ${\rm Z/X}=0.0231 \pm 0.005$, when calibrating the models\footnote{Although recent analyses using 3D hydrodynamical models of the solar atmosphere \citep{Asplund_etal2005, Lodders_etal2009, Caffau_etal2011} have resulted in lower values for Z/X compared to the previous GS98 mix, we do not use these lower values because they disagree with high-precision helioseismological determinations of the solar structure \citep[e.g.,][ and references therein]{Basu_Antia2004, Serenelli_etal2009, Serenelli_etal2011, Bi_etal2011, Antia_Basu2011}.}. Using this solar composition, we computed a Standard Solar Model, assuming an age for the Sun of 4.6 Gyr. By fitting the present solar luminosity, radius and Z/X ratio, we obtained a mixing length parameter $\alpha=1.896$, an initial solar helium abundance ${\rm Y}_\odot=0.262$ and a global solar metallicity ${\rm Z}_\odot=0.0167$.

\subsection{Isochrones}

For studying GCs, if the canonical SSP is accepted, it can be assumed that all stars were formed at the same time with the same chemical composition. Since higher-mass stars evolve more rapidly, one cannot compare the evolutionary tracks for an SSP directly with the GC observations. Instead, it is necessary to build isochrones which connect the points at the same age along the evolutionary tracks with different masses.

To construct our isochrones, we use the equivalent evolutionary points (EEP) technique first proposed by \citet{Prather1976}. These EEPs represent points with similar properties (which can be internal or external) for tracks of different masses. For low-mass stars, the EEPs are:

\begin{figure}[t]
 \centering
    \includegraphics[height=9cm]{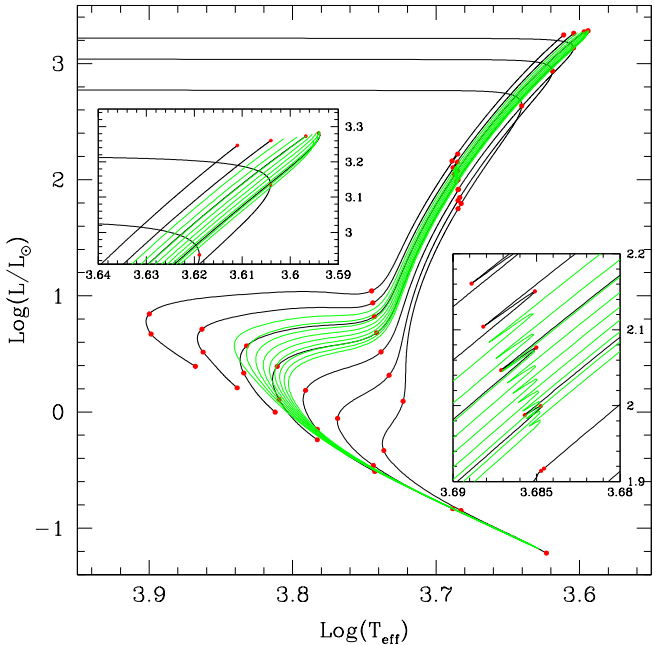}
    \caption{Set of evolutionary tracks (black lines) for stars between 1.1 (hottest track) and 0.5 $M_\odot$ (coolest track) with Y=0.245, Z=9.3$\times$10$^{-4}$, and [$\alpha$/Fe]=+0.3, plotted at $\Delta M = 0.1 M_\odot$ intervals. Red points are the EEPs for each track, while green lines are isochrones, from left to right, between 7 and 15 Gyr, in steps of 1 Gyr. Inner panels are zoom-ins to the RGB tip (left) and RGB bump (right).}
    \label{IsochroneMaker}
\end{figure}

\begin{itemize}
\item {\bf EEP1:} This point is defined as the beginning of the MS, where we take, as a first approximation for low-mass stars, the age to be zero (for SECs including the pre-MS evolution, this point is defined when the luminosity due to gravitational collapse decreases to 0.01 $L_\odot$). 

\item {\bf EEP2:} This point is defined when the helium abundance in the center reaches a value of 90\%, which is near the point when low-mass MS stars have reached their maximum effective temperature.

\item {\bf EEP3:} This point is similar to EEP2, but is determined when the central hydrogen is completely depleted.

\item {\bf EEP4:} This is the point where the mass of the convective envelope is 25\% of the total mass, which characterizes the base of the RGB.

\item {\bf EEP5:} This point is determined when the star's luminosity starts to decrease as the H burning shell arrives at the chemical composition 
discontinuity previously produced when the envelope convection zone reached its maximum depth.

\item {\bf EEP6:} This is where the star's luminosity starts to increase again, as the stellar structure readjusts after EEP5 to the higher H abundance outside the composition discontinuity. For the cases where the RGB bump does not exist ($M \lesssim 0.7M_\odot$ in Fig. \ref{IsochroneMaker}), EEP5 and EEP6 are determined by extrapolating the RGB bump positions for lower-mass stellar tracks with well-defined RGB bumps. 

\item {\bf EEP7:} Finally, the last point is defined when $\log (L_{\rm He}/L_\odot) >$ 2.0 due to the beginning of the helium flash or when $T_{\rm eff}$ starts to increase due to the star's leaving the RGB before igniting He in the core.

\end{itemize}

The definition for these points was kindly provided by S. Yi (2008, private communication), except for EEP3 which was added to obtain a better resolution, and EEP7 where the original definition took into account only the luminosity via helium burning. Indeed, the original definition of EEP7 was valid only if the star could ignite helium, whereas our new definition is valid for any star \citep[see also][]{Bergbusch_Vandenberg1992}. Figure \ref{IsochroneMaker} shows an example of some isochrones between 7 and 15 Gyr, which were created using PGPUC evolutionary tracks, the technique presented before, and the Hermite interpolation algorithm for constructing reasonable analytic curves through discrete data points presented by \citet{Hill1982}. 

Finally, we emphasize that the EEP technique can also be applied to tracks for different Y, Z, [$\alpha$/Fe], and any other parameter (mixing length parameter, mass loss rates, overshooting, etc.). This means that one can always create new evolutionary tracks with a parameter inside the original set of tracks, maintaining the others constants. For example, instead of interpolating in mass (as was used to create an isochrone), one can interpolate in Y to create a new evolutionary track for a different helium abundance while maintaining fixed the other parameters (initial mass, Z, [$\alpha$/Fe], C+N+O abundance, etc.). This procedure can be repeated for each parameter, with the end result that one can always create a new evolutionary track (and consequently isochrones) with any parameter inside the grid of original parameters. Such a procedure is implemented in the PGPUC Online web page (see Appendix \ref{AppPGPUConline}).

\subsection{ZAHB loci}
\label{ZAHBloci}

At the moment when the helium flash and the subsequent secondary flashes have completely removed the degeneracy in the helium core, the star has arrived at the ZAHB. In this phase, the star is burning helium in the core, while in an outer shell it keeps burning hydrogen. Because stars in the HB phase spend $\sim$ 50\% of their life without varying their luminosity by a large fraction ($\lesssim$ 5\%, based on PGPUC calculations), it is useful to create ZAHB loci which represent the lower boundary of the HB loci.

\citet{Serenelli_Weiss2005} have compared different methods to construct ZAHB models without computing the complete evolution for several stars with different masses and mass loss rates. They concluded that ZAHB loci could be well-defined if a star is evolved without mass loss from the MS to the HB (passing through the flash phase), and then stopping the temporal evolution of this star at the point when the rate of change in the internal thermal energy is smallest.  At this point the carbon abundance in the center is 5\%, and the core convection has reached the center. This defines the red end of the ZAHB sequence. Then mass from the envelope is removed until the minimum envelope mass is reached, which corresponds to the removal of almost all of the envelope at the hottest point along the ZAHB sequence. The ZAHB locus corresponds to the line connecting all these ZAHB models, computed for different envelope masses but with the same core mass. We will therefore adopt the method suggested by \citet{Serenelli_Weiss2005} when constructing ZAHB loci with the PGPUC SEC.

\section{Effects of $\Delta$Y}
\label{EffectsHe}

As is pointed out in \S\ref{intro}, a spread in helium abundance among stars has been suggested to explain why some GCs show multiple populations in their CMDs, which is difficult to verify because the initial abundance of this element can only be measured, in such old populations, at high resolution and with high S/N, over a small range in temperatures along the HB \citep{Villanova_etal2009,Villanova_etal2012}. In this section, we show the effects of helium enrichment upon evolutionary tracks, isochrones and ZAHB loci using PGPUC models, which give us some clues of what we should observe in GC's color-magnitude diagrams when a helium enhancement is present.

\subsection{Effects of $\Delta$Y on [Fe/H]}

In Table \ref{TableChemical} we show how [Fe/H] increases with increasing Y for a constant Z. Thus if multiple populations in a GC are only formed with enriched helium and without any enrichment in metals \citep[e.g.,][]{Valcarce_Catelan2011}, stars with different Y will show different [Fe/H] values \citep[see also][]{Sweigart1997b,Bragaglia_etal2010b}. For example, if in a GC a population exists with a primordial initial helium abundance (Y$_{\rm FG}$=0.245) alongside another population with Y$_{\rm SG}$=0.345, and if both populations have the same metallicity Z =$1.6\times 10^{-3}$ (corresponding to [Fe/H] = -1.25 when Y = 0.245), then one should observe a spread in the iron abundance by $\Delta$[Fe/H]$\sim$0.063. In fact, the analytical formula for this spread in [Fe/H] is given straightforwardly by

\begin{equation}
\label{EqIron}
{\Delta \rm [Fe/H]} = {\rm [Fe/H]_{FG}-[Fe/H]_{SG}}= \log \left( \frac{\rm 1-Y_{SG}-Z}{\rm 1-Y_{FG}-Z} \right),
\end{equation}

\noindent where the subscripts FG and SG refer to the first and second generations, respectively.

From spectroscopic observations \citet{Bragaglia_etal2010} have found a spread of $\Delta$[Fe/H]=0.070 among the RGB stars in NGC~2808 \citep[average $[{\rm Fe/H}\rbracket$=-1.15,][]{Carretta_etal2009a}, implying a spread in helium of $\Delta$Y$\sim$0.1 from equation \ref{EqIron} for a metallicity of Z=$2.18 \times 10^{-3}$. This helium spread is close to the value suggested to reproduce the difference between the bluest and reddest MSs in this particular GC \citep{DAntona_etal2005, Piotto_etal2007}. 

On the other hand, in addition to the already discussed possibility of a variation in the CNO abundances, and thus different metallicity at fixed [Fe/H] for the enhanced populations, there is a small possibility that second generation stars could have higher Y and lower Z than the first generation, if there was capture of metal-poor material from outside of the GC which was then mixed with material processed inside the GC \citep{Conroy_Spergel2011}.

\subsection{Effects of $\Delta$Y on evolutionary tracks}
\label{EffectsHeETs}

\begin{figure}
  \centering
    \includegraphics[height=9cm]{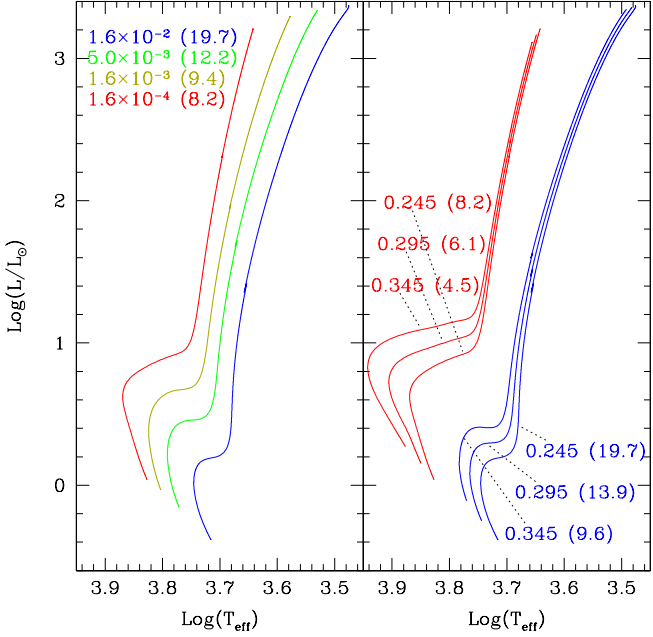}
    \caption{Evolutionary tracks for stars with 0.9 $M_\odot$ from the MS to the tip of the RGB. The left panel shows the  effects produced by different metallicities (from left to right: Z=$1.6\times 10^{-4}$, $1.6\times 10^{-3}$, $5.03\times 10^{-3}$, and $1.57\times 10^{-2}$) with Y=0.245. The effects produced by different helium abundances (Y=$0.245$, $0.295$, and $0.345$) for two metallicities ([Fe/H]=$-2.25$, and $-0.25$) are shown in the right panel. In parentheses are labeled the age in Gyr at the RGB tip for each evolutionary track.}
  \label{FigTrackDifZ}
\end{figure}

Before comparing isochrones and ZAHB loci, we show the effects of changing the metallicity and the helium abundance on the evolutionary tracks for stars with 0.9 $M_\odot$ (see Fig. \ref{FigTrackDifZ}). Even though many of these effects have been known since the first evolutionary codes appeared \citep[][]{Simoda_Iben1968, Aizenman_etal1969, Iben1974, Simoda_Iben1970}, we summarize them again using PGPUC stellar models. Although the input physical ingredients have been updated over the years, the main effects are almost the same.

 \begin{figure}[t]
   \centering
      \includegraphics[height=9cm]{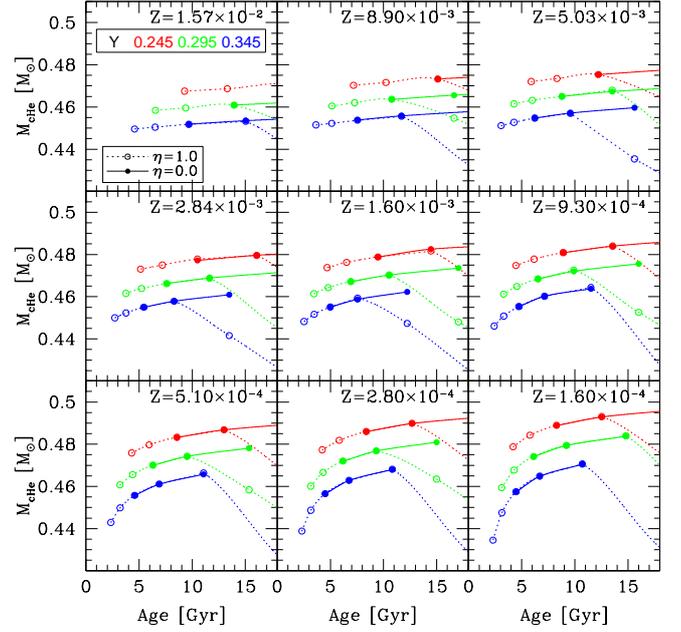}
      \caption{Relation between age at the RGB tip and the He core mass ($M_{\rm cHe}$) reached for stars with different initial masses, metallicities, helium abundances and different mass loss rates (solid lines: no mass loss, or $\eta = 0$; dotted line: $\eta = 1.0$). Interpolated lines connect the real values for different initial helium abundances Y, from top to bottom: $0.245$ (red), $0.295$ (green), and $0.345$ (blue). For stars with $\eta = 1.0$ and $\eta = 0.0$, the initial masses go, from left to right, from $1.1$ to $0.5\, M_{\odot}$ and from $0.9$ to $0.7\, M_{\odot}$, respectively, in steps of $0.1\, M_{\odot}$. As can be seen from the monotonic behavior of the solid lines, the helium core mass increases for higher ages (or lower initial masses). In the case of stars with mass loss (dotted lines), there is a maximum value for $M_{\rm cHe}$ because, for the mass loss efficiency assumed ($\eta = 1.0$), stars lose a large fraction of their masses at higher ages, which decreases the hydrogen-burning rate and leads to a premature departure from the RGB; a hot flasher may then take place. (See text for further details.) }
      \label{FigMcTip}
 \end{figure}

As is observed in the left panel of Fig. \ref{FigTrackDifZ}, stars with lower metallicities have higher luminosities and effective temperatures from the ZAMS to the tip of the RGB. This occurrence is mainly due to the decrease of the radiative opacity when the metallicity decreases, thus enhancing the outward transport of energy from the core to the surface. As a result, metal-poor stars have a denser and hotter core, which increases the H-burning nuclear rates. This is also the reason for which the core H-burning lifetime decreases with decreasing metallicity. 

Moreover, the temperature of the RGB decreases faster for high metallicities due to the higher opacity and the consequent faster increase of the radius. At the RGB bump, two additional effects are observed: i)~its luminosity decreases for higher metallicities due to a deeper penetration of the convective envelope into the interior during the first dredge-up and ii)~its prominence also increases for higher metallicities due to the larger discontinuity in the H abundance, and hence in the radiative opacity, at the deepest point reached by the convective envelope \citep{Thomas1967, Iben1968, Sweigart1978, Sweigart_Gross1978, Lee1978, Yun_Lee1979}.

Finally, we note that stars with a higher metallicity reach a higher luminosity at the tip of the RGB \citep{Iben_Faulkner1968, Demarque_Mengel1973}. This result is a consequence of two competing effects. First, the energy output of the hydrogen-burning shell in a more metal-rich star on the RGB is greater at each value of the He core mass due to the greater abundance of the CNO elements which increases the nuclear reaction rates of the CNO cycle. As a consequence the He core then grows and heats up more rapidly, thereby causing He ignition to occur at a lower He core mass \citep{Sweigart_Gross1978, Cavallo_etal1998}. However, the greater hydrogen-shell luminosity more than compensates for the smaller core mass at helium ignition, leading to the increased RGB tip luminosity with metallicity shown in Figure \ref{FigTrackDifZ}.

\begin{figure*}
   \centering
      \includegraphics[height=18.0cm, trim=0cm 1cm 0cm 0cm]{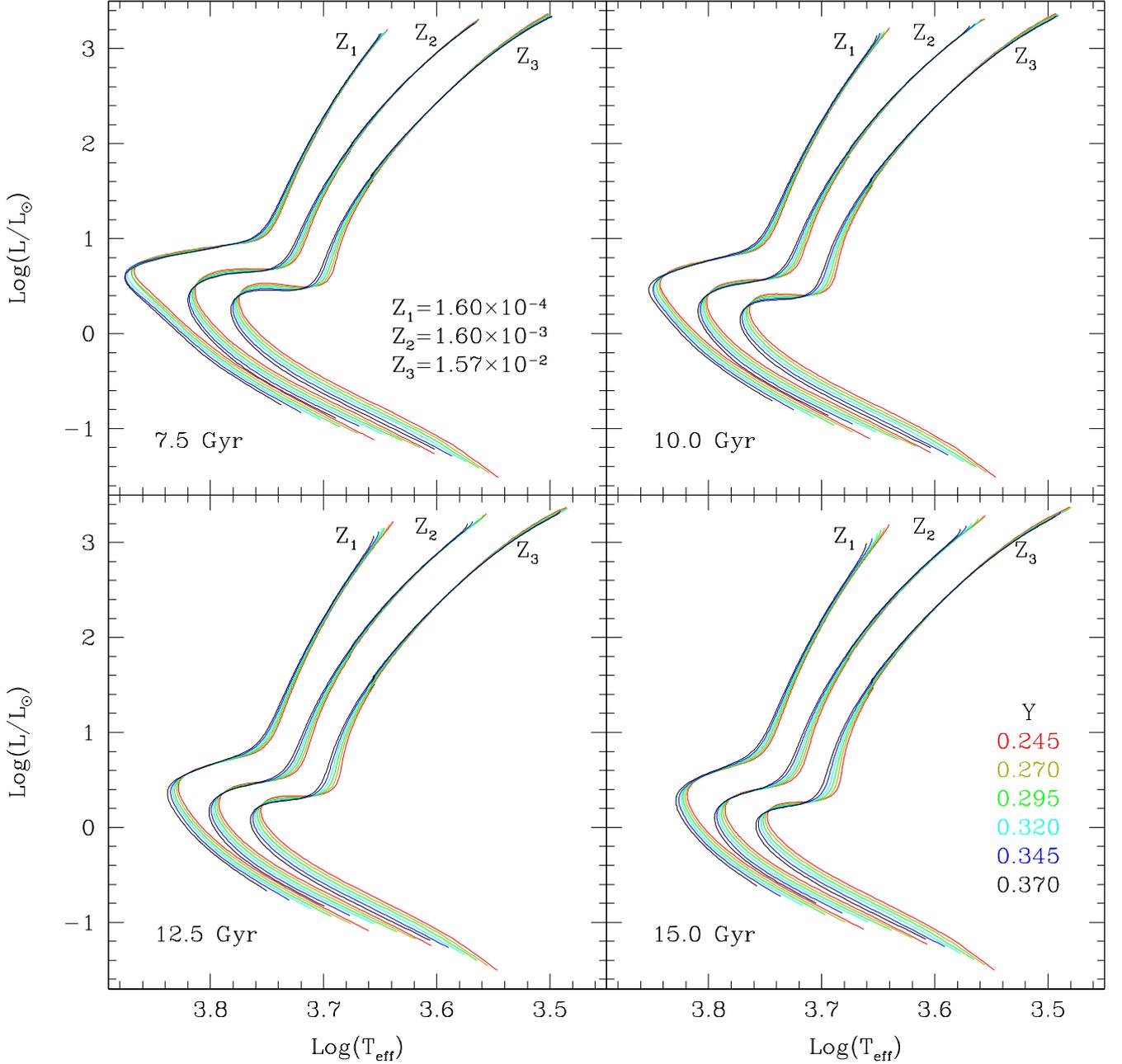}
      \caption{Isochrones for (from upper left to lower right) 7.5, 10.0, 12.5, and 15.0 Gyr with metallicities, from left to right, Z$_1$=$1.60 \times 10^{-4}$, Z$_2$=$1.60\times 10^{-3}$, and Z$_3$=$1.57\times 10^{-2}$ in the $\log L$-$\log T_{\rm eff}$ plane. Different colors represent different initial helium abundances: Y=0.245 (red), 0.270 (yellow), 0.295 (green), 0.320 (cyan), 0.345 (blue), and 0.370 (black).}
      \label{FigIsochrones}
 \end{figure*}

In the case of different helium abundances (right panel of Fig. \ref{FigTrackDifZ}), helium--rich ZAMS stars have higher luminosity and $T_{\rm eff}$ \citep{Demarque1967, Iben_Faulkner1968, Simoda_Iben1968, Simoda_Iben1970, Demarque_etal1971, Hartwick_vandenBerg1973, Sweigart_Gross1978}. To maintain adequate pressure, i.e., hydrostatic equilibrium, a helium-rich star must have a higher temperature in the center to compensate for the fewer particles which have to support the same mass, with the end result that these stars are more compact, which increases the $T_{\rm eff}$. In addition, the opacity is also lower, and then the greater energy being produced in the center can more easily flow outward when the helium abundance is increased.

Even though stars with different Y arrive on the RGB with different luminosities, the slope of the track along the RGB in a $\log L$--$\log T_{\rm eff}$ diagram is very similar. However, there are an increase of the luminosity and a decrease of the extension of the RGB bump when Y is increased, due to the fact that the convective envelope in a helium-rich star does not penetrate as deeply during the first dredge-up and therefore the composition discontinuity produced at the deepest penetration point is then both smaller and further out in mass.

Finally, the luminosity of the RGB tip decreases when Y is increased because the flash occurs at a smaller core mass (see Fig. \ref{FigMcTip}). This is a consequence of the faster growth in the core mass and hence faster increase in the maximum core temperature.

\subsection{Effects of $\Delta$Y on isochrones}

\begin{figure}
   \centering
      \includegraphics[width=9.3cm, height=8.75cm]{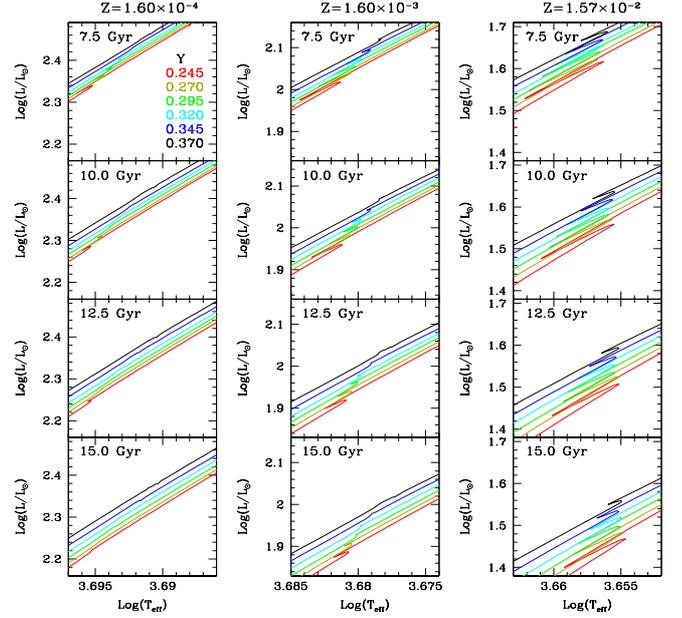}
      \caption{Isochrones for the labeled metallicities at the RGB bump. Colors represent the same Y values as in Fig. \ref{FigIsochrones}.}
      \label{FigBumpIsochrones}
 \end{figure}

Figure \ref{FigIsochrones} shows isochrones with ages between 7.5 and 15.0 Gyr, for several Z values and helium abundances from Y=0.245 to 0.370, whereas Figure \ref{FigBumpIsochrones} shows in more detail the RGB bump. As can be observed in these figures, there are several effects on isochrones when Y is increased whose detailed study would be quite time-consuming. For this reason, here we summarize them qualitatively, to give us hints of what places in HR diagrams we should focus our attention to, when we tabulate these effects quantitatively; a preliminary discussion can also be found in \citet{Catelan_etal2010}. Again, we emphasize that several of the following effects were directly or indirectly found in the past \citep[e.g.][]{Demarque1967, Iben_Faulkner1968, Simoda_Iben1968, Simoda_Iben1970, Demarque_etal1971, Hartwick_vandenBerg1973, Sweigart_Gross1978, Lee1978, Yun_Lee1979}.

\begin{itemize}
\item {\bf MS:} As was pointed out for the case of evolutionary tracks, when Y is increased the effects on MS stars with equal masses include an increase in both $L$ and $T_{\rm eff}$. Then, if the difference in $L$ between isochrones of the same age is compared for the same $T_{\rm eff}$, this is a comparison between stars with different masses, where the more massive stars are the ones with lower Y. Finally, since low-mass stars and/or stars with low He abundance have longer evolutionary times, the size of the difference in luminosity between MS stars with different Y but the same Z almost does not depend on age in the range of ages studied here.

\item {\bf TO:} When isochrones for the same Z are compared, the TO point is affected by the difference in Y and also by the age. While an increase in Y drives a reduction of $L$ and an increase in $T_{\rm eff}$, an increase in age decreases both $L$ and $T_{\rm eff}$. As can be seen, at the TO point the difference in $\log T_{\rm eff}$ between isochrones with different Y changes with age at almost the same rate independently of the metallicity, but the difference in $\log (L/L_\odot)$ changes faster with age for higher metallicities.

\item {\bf SGB:} Here, the difference produced by an increase in Y is observed as a decrease in $L$. However, this difference is more prominent for low ages or high metallicities \citep[see also ][]{DiCriscienzo_etal2010}, or in other words, when the difference in the initial masses of the stars in this zone is large, the difference in the $\log L$-$\log T_{\rm eff}$ loci is also large.

\item {\bf Lower RGB:} In the lower RGB there are three important effects to consider when Y is increased: i)~the minimum luminosity almost does not depend on Y, but depends on Z and age; ii)~as in the TO point, $T_{\rm eff}$ at the base of the RGB is higher for higher helium abundances, while the difference in $T_{\rm eff}$ between isochrones with different Y increases for higher ages; iii)~finally, the absolute value of the lower RGB slope is higher for lower helium abundances.

\item {\bf RGB bump:} The effect of an increase in Y at the RGB bump is shown in Fig. \ref{FigBumpIsochrones}, where an increase of $L$ and a reduction of the extension of the RGB bump can be seen (see Sect.\ref{EffectsHeETs}). However, the maximum $T_{\rm eff}$ of the RGB bump decreases for higher Y, but this effect is more evident for lower metallicities. An increase in age also increases the difference in $L$ between Y values, while the difference in $T_{\rm eff}$ almost does not change with age. These effects may be observed as changes of the shape of the RGB bump in luminosity functions, where for large extensions of the theoretical RGB bump the shape is bimodal, while for small extensions the shape is single-peaked \citep{Cassisi_etal2002}.

\item {\bf Upper RGB:} Above the RGB bump, as was pointed out in Sect.\ref{EffectsHeETs}, the difference in Y does not induce any great difference in the RGB slope, the only observable effect being a shift in the position of the RGB tip, whose luminosity decreases for higher helium abundances. However, this reduction of the RGB tip luminosity (and the luminosity function at this position) should be very difficult to observe in color-magnitude diagrams of GCs, due to the small number of stars near this point.
\end{itemize}

\subsection{Effects of $\Delta$Y on ZAHBs}

\begin{figure*}
   \centering
      \includegraphics[width=19cm, trim=0.7cm 13.5cm 0cm 0cm]{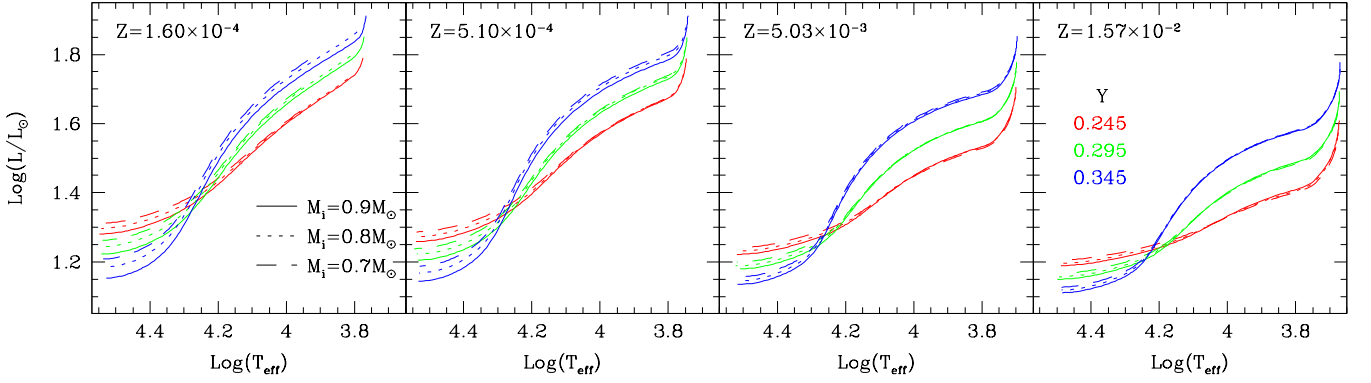}
      \caption{ZAHB loci for different metallicities (increasing from upper left to bottom right) and initial helium abundances. For each Y value, ZAHB loci were constructed using three RGB progenitors with different initial masses (M$_i$), which can be related to an age dependence of the ZAHB, which is particularly evident at low metallicities.}
      \label{FigZAHBdifZ}
 \end{figure*}

As for the isochrones, in several previous works \citep[e.g.,][]{Faulkner_Iben1966, Iben_Faulkner1968, Iben_Rood1970, Sweigart1987} the effects of different helium abundances have been studied in ZAHB loci and HB models for different Z values. Here, we summarize the main effects for ZAHB loci.

Figure \ref{FigZAHBdifZ} shows changes in the ZAHB locus when the He abundance is increased for several metallicities. When Z is constant, it is observed that for high Y i)~cooler HB stars (the more massive ones) have high ZAHB luminosities, and ii)~hotter HB stars (the less massive ones) have lower ZAHB luminosities. Cooler HB stars show this effect for the same reason MS stars are brighter: high helium abundances decrease the radiative opacity, thus stars are more compressed (increasing the internal temperature and density), leading to an increase in the hydrogen burning rate, while the helium burning rate at the core decreases due to the smaller helium core mass (see Fig. \ref{FigMcTip}), but the luminosity's decrease due to the helium burning is lower than the luminosity's increase due to the hydrogen burning, with the end result that the total luminosity increases. In the case of hotter HB stars, in view of the small He core, together with the fact that hotter HB stars have tiny envelopes and thus small hydrogen burning rates, these stars have lower luminosities when Y is higher: their luminosity comes chiefly from the helium burning rate in the (smaller) core.

Moreover, the luminosity difference between ZAHBs with different Y depends on the metallicity, where this difference between cooler HB stars is smaller for lower Z, but the difference between hotter HB stars is higher for lower Z. The higher increase in luminosity for higher metallicity when Y is increased is due to the increase of the efficiency of the CNO cycle (due to the higher Z) that makes stronger the effect of the increase of Y mentioned in the previous paragraph. In the case of hotter HB stars, the difference in luminosity is due to the difference in the helium core masses, which are greater for lower metallicities (Fig. \ref{FigMcTip}).

Although \citet{Caputo_deglInnocenti1995} have shown that the luminosity of the ZAHB at a fixed $T_{\rm eff}$ depends on age for low metallicities, this effect has not been (recently) studied in detail for the complete range of $T_{\rm eff}$ and for different Y. In the next section we show the effects of age upon ZAHB loci, which must be taken into account for all metallicities.

\subsubsection{ZAHB dependence on age}
\label{ZAHBage}

\begin{table}[!b]
\center
\begin{tabular}{ccc|rrrr|}
\hline
\multicolumn{3}{|c}{Z=$1.60\times 10^{-4}$ }             & \multicolumn{4}{|c|}{$\log T_{\rm eff}$} \\
\hline
\multicolumn{1}{|c}{Y}     &$\Delta M_i$& $\Delta$ Age & 4.45 & 4.20 & 4.00 & 3.83 \\
\hline
\multicolumn{1}{|c}{0.245} & 0.9-0.8    &   -4.2       & 0.040& 0.023& 0.010& 0.010\\
\multicolumn{1}{|c}{0.245} & 0.9-0.7    &  -11.9       & 0.075& 0.045& 0.015& --   \\
\multicolumn{1}{|c}{0.295} & 0.9-0.8    &   -3.0       & 0.055& 0.033& 0.025& 0.028\\
\multicolumn{1}{|c}{0.295} & 0.9-0.7    &   -8.7       & 0.100& 0.055& 0.038& --   \\
\multicolumn{1}{|c}{0.345} & 0.9-0.8    &   -2.2       & 0.080& 0.053& 0.048& 0.053\\
\multicolumn{1}{|c}{0.345} & 0.9-0.7    &   -6.2       & 0.143& 0.083& 0.078& --   \\
\hline
\hline
\multicolumn{3}{|c}{Z=$1.60\times 10^{-3}$}             & \multicolumn{4}{|c|}{$\log T_{\rm eff}$} \\
\hline
\multicolumn{1}{|c}{Y}     &$\Delta M_i$& $\Delta$ Age & 4.45 & 4.20 & 4.00 & 3.83 \\
\hline
\multicolumn{1}{|c}{0.245} & 0.9-0.8    &   -5.0       & 0.030& 0.015& 0.003& 0.003\\
\multicolumn{1}{|c}{0.245} & 0.9-0.7    &  -14.1       & 0.058& 0.030& 0.000&-0.003\\
\multicolumn{1}{|c}{0.295} & 0.9-0.8    &   -3.6       & 0.033& 0.018& 0.010& 0.010\\
\multicolumn{1}{|c}{0.295} & 0.9-0.7    &  -10.1       & 0.068& 0.028& 0.015& 0.013\\
\multicolumn{1}{|c}{0.345} & 0.9-0.8    &   -2.6       & 0.040& 0.023& 0.020& 0.020\\
\multicolumn{1}{|c}{0.345} & 0.9-0.7    &   -7.2       & 0.078& 0.038& 0.033& 0.030\\
\hline
\hline
\multicolumn{3}{|c}{Z=$1.57\times 10^{-2}$}             & \multicolumn{4}{|c|}{$\log T_{\rm eff}$} \\
\hline
\multicolumn{1}{|c}{Y}     &$\Delta M_i$& $\Delta$ Age & 4.45 & 4.20 & 4.00 & 3.83 \\
\hline
\multicolumn{1}{|c}{0.245} & 0.9-0.8    &  -10.7       & 0.018& 0.013&-0.003&−0.010\\
\multicolumn{1}{|c}{0.245} & 0.9-0.7    &  -30.1       & 0.040& 0.030&-0.005&−0.020\\
\multicolumn{1}{|c}{0.295} & 0.9-0.8    &   -7.6       & 0.018& 0.010&-0.003&-0.003\\
\multicolumn{1}{|c}{0.295} & 0.9-0.7    &  -21.4       & 0.043& 0.025&-0.005&-0.010\\
\multicolumn{1}{|c}{0.345} & 0.9-0.8    &   -5.3       & 0.020& 0.008& 0.005& 0.005\\
\multicolumn{1}{|c}{0.345} & 0.9-0.7    &  -15.0       & 0.043& 0.010& 0.003& 0.005\\
\hline
   \end{tabular}
	\caption{Difference in the bolometric magnitude $m_{\rm bol}$ (four last columns) between ZAHB loci with different progenitor masses ($\Delta M_i$), for 3 metallicities and 3 helium abundances (first column) at the labeled $\log T_{\rm eff}$. In the third column are listed the difference in the ages (in Gyr) at the ZAHB of the progenitor stars used to calculate the difference in $m_{\rm bol}$.}
	\label{TableZAHB1}
\end{table}

In this section we stress the ZAHB loci dependence on age, and how this effect is more important at low metallicity and/or high helium abundances. Although this effect has been known for a long time \citep[e.g.,][]{Iben1974}, so far it has been largely neglected. However, nowadays, due to the high precision of the available photometric and spectroscopic data for HB stars \citep[see, e.g.,][]{Catelan_etal2009, Dalessandro_etal2011, Moehler_etal2011, MoniBidin_etal2011b}, it is increasingly important to determine the ZAHB locus for the appropriate He core mass, which (as already discussed in Iben's review) depends on the metallicity, initial helium abundance {\em and} the RGB progenitor mass (and thus the age). This is especially true when blue HB, extreme blue HB and blue hook stars are studied, and even more so when the bluest filters are used.

In regard to the time required for stars to reach the RGB tip ($t_{\rm Tip}$), it depends on three main factors: i)~the initial helium abundance, where $t_{\rm Tip}$ decreases with higher Y because the H burning rate increases; ii)~ the heavy element abundance, where $t_{\rm Tip}$ increases with higher Z because density and temperature in the stellar interior are lower in metal-rich stars, which decreases the H burning rate at the MS since the burning rate by the CNO cycle is negligible; iii)~the initial mass, since more massive stars have higher internal temperatures, thus $t_{\rm Tip}$ decreases \citep{Iben_Rood1970,Hartwick_vandenBerg1973}. These effects are shown in Figure \ref{FigMcTip}, for different [Fe/H], Y, and initial masses.

As Figure \ref{FigZAHBdifZ} shows, the progenitor mass used to create ZAHB loci changes the ZAHB luminosity due to the different helium core masses reached by these stars at the RGB tip. The helium core mass at the RGB tip is determined by a competition between the hydrogen burning luminosity which controls the rate at which the helium core grows and heats up and the cooling rate via thermal conduction and neutrino emission during the RGB evolution. To produce the helium flash, $T \sim 10^8$ K and $\rho \sim 10^6$ g/cm$^{3}$ are required as condition, which is reached when some fraction of the helium core is completely supported by degenerate electron pressure.

As is shown in Fig. \ref{FigMcTip}, the parameters that determine the final helium core mass are the helium abundance, initial mass, and amount of mass loss on the RGB~-- the latter two quantities depending on the age (for a given chemical composition). The sense of the dependence of $M_{\rm cHe}$ on Y and initial mass is that an increase in either increases the temperature at the hydrogen-burning shell, and consequently the helium flash takes place earlier and with a lower $M_{\rm cHe}$. The dependence on mass loss becomes relevant only when the total RGB mass loss becomes quite substantial, which may lead to a decrease in efficiency of the H-burning shell, and hence to a He flash that may take place only after the star has already left the RGB, while the star is on its way to the white dwarf cooling curve, or even on the white dwarf cooling curve proper \citep[e.g.,][]{Castellani_Castellani1993, DCruz_etal1996, Brown_etal2001, Moehler_etal2011}. This effect explains the downward variation of $M_{\rm cHe}$ with age towards old ages that is seen in Fig.~7, particularly for ${\rm Y}>0.245$ and $\eta=1.0$. The metallicity also plays an important r\^ole in the final $M_{\rm cHe}$ value, but an increase in metallicity induces two effects: i) a decrease in the temperature at the hydrogen-burning shell; and ii) an increase of the CNO cycle efficiency. Then, which one of these effects is more important depends on the initial mass of the star, as observed in the change of the slope of $M_{\rm cHe}$ with $Z$ in Figure \ref{FigMcTip}. The temperature of the hydrogen burning shell is determined solely by the mass of the helium core. This is the same reason why an RGB star evolves to higher luminosities. As in a white dwarf, the higher the mass of the helium core, the smaller its radius, but the temperature of the outer layer of the core is higher, which increases the temperature of the hydrogen shell around the core with the consequent increase of the hydrogen burning rate \citep{Renzini_etal1992}.

 \begin{figure}
   \centering
      \includegraphics[height=7.5cm, trim=0.0cm 0.5cm 0cm 0cm]{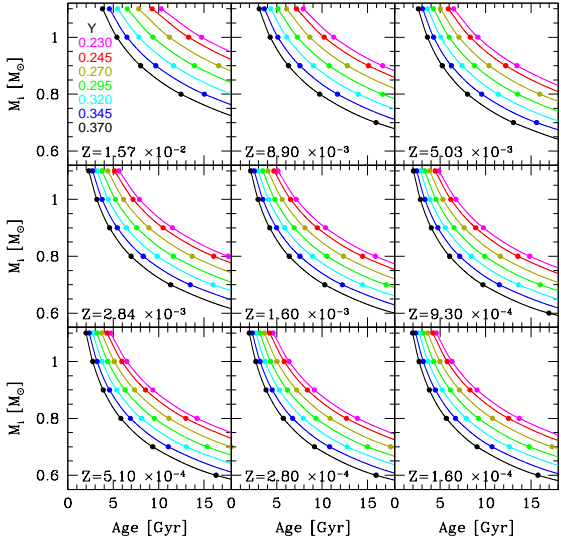}
      \caption{Relation between age at the RGB tip and the initial stellar mass ($M_i$) depending on the initial helium abundance and the metallicity. Colors represent different helium abundances. Dots, and lines have the same meaning as in Fig. \ref{FigMcTip}. }
      \label{FigMiTip}
 \end{figure}

 \begin{figure}[!h]
   \centering
      \includegraphics[height=12.4cm, trim=0.0cm 1.0cm 8cm 0cm]{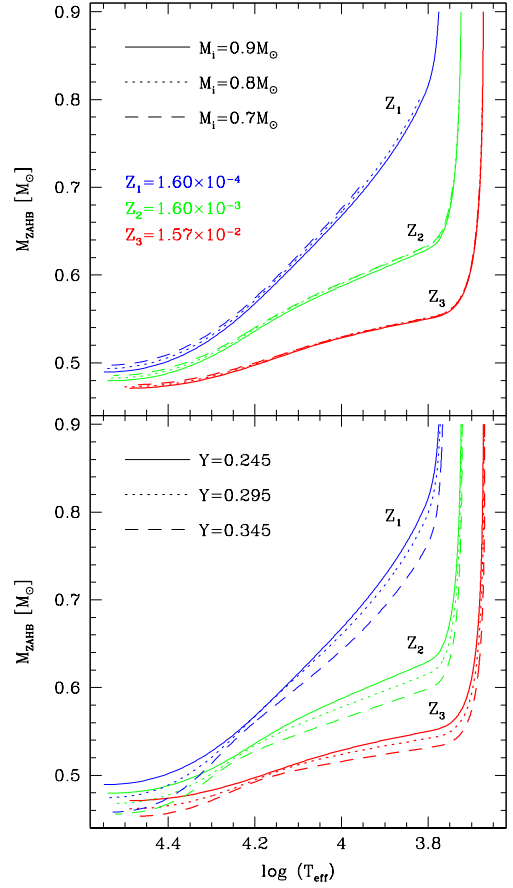}
      \caption{Dependence of the effective temperature of ZAHB stars on their mass for three different metallicities: $Z_1=1.60\times 10^{-4}$, $Z_2=1.60\times 10^{-3}$ and $Z_3=1.57\times 10^{-2}$. }
      \label{FigZAHB.MvsT}
 \end{figure}

 \begin{figure*}[t]
   \centering
      \includegraphics[height=7cm, trim=0.6cm 12.0cm 0cm 0.7cm]{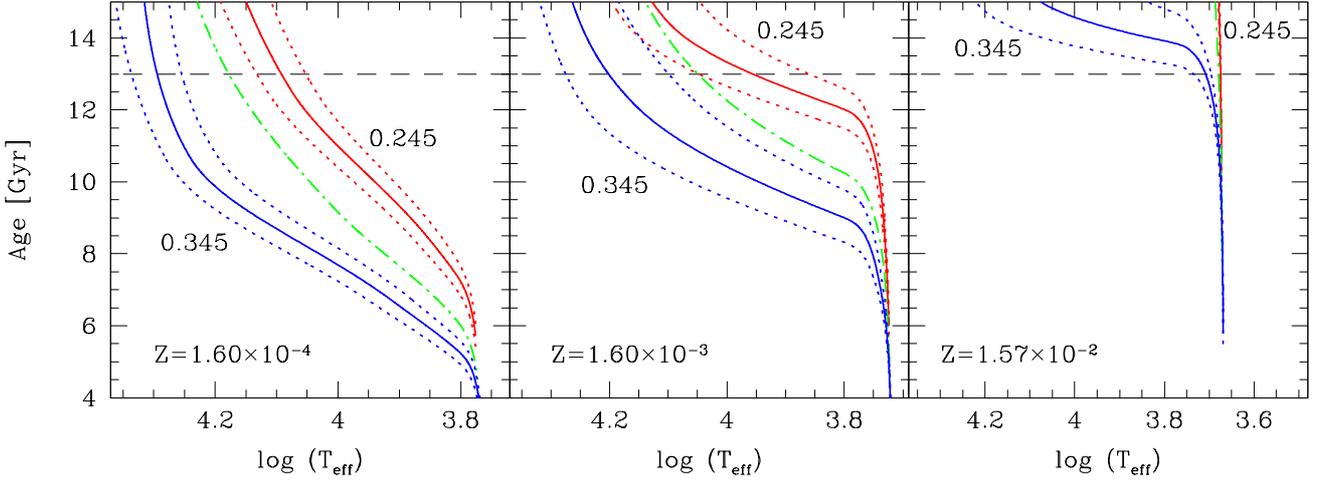}
      \caption{Effective temperature of the stars arriving to the ZAHB for ages between 4 and 15 Gyr for 3 initial helium abundances Y$=0.245$ (coolest continuous line), Y$=0.295$ (dotted-dashed line) and Y$=0.345$ (hottest continuous line), and 3 metallicities: $1.60\times 10^{-4}$ (left panel), $1.60\times 10^{-3}$ (middle panel), and $1.57\times 10^{-2}$ (right panel). Dotted lines show the change in the effective temperatures when a dispersion of $0.02 M_\odot$ in the total mass loss is considered. Dashed line shows a age of 13 Gyr.}
      \label{FigZAHB.TvsAge}
 \end{figure*}

Depending on the helium core mass, the luminosity of the ZAHB will change. Figure \ref{FigZAHBdifZ} shows this effect in HR diagrams, where ZAHB loci can be misrepresented if the mass used to construct them does not give the required helium core mass for the stellar population being studied. For example, ZAHB loci for a GC with [Fe/H]=-1.48 and three populations with Y=0.245, 0.295, and 0.345 at an age of 12.5 Gyr, must be created from RGB progenitor masses of $\sim$0.83, $\sim$0.75, and $\sim$0.68 $M_\odot$, respectively. However, if the age is 7.5 Gyr, the RGB progenitor masses for these ZAHBs must be $\sim$0.95, $\sim$0.87, and $\sim$0.79 $M_\odot$. Then, when ZAHB loci are used together with isochrones, the initial mass used to create the ZAHB must be defined by the initial mass of the star at the RGB tip of the respective isochrone, which depends on Y and Z (see Fig. \ref{FigMiTip}). This means that the whole ZAHB depends on the age of the GC. This property is more important for high helium abundance and/or lower metallicities 

In Table \ref{TableZAHB1} are shown the errors in bolometric magnitude $m_{\rm bol}$ that can be induced if ZAHB loci were created with a progenitor mass of 0.9 $M_\odot$ instead of a progenitor mass determined according to the age of the GC. When GCs with multiple populations are studied, the mass of the progenitor mass of the ZAHB has to be defined by Fig. \ref{FigMiTip} if it is assumed that all the stellar populations were born at almost the same time. Otherwise, the mass of the progenitor of the ZAHB for the younger population has to be defined by Fig. \ref{FigMiTip} but with $M_i$-Age relation of the given helium abundance displaced in the time that this population required to be formed.

\subsubsection{ZAHB morphology in multiple populations}
\label{ZAHBEffects}

The HB morphology of GCs has been studied for several decades, but without a consensus about the second parameter (together with the metallicity) to parameterize the HB morphology (see section \ref{intro}). Historically, multimodality on the HB has been used as an argument in favor of the importance of other second parameter candidates, besides cluster age \citep[e.g.,][see also \citealt{Catelan2008} for a recent review on the subject of gaps and multimodality on the HB]{Norris1981, Norris_Smith1983,Crocker_etal1988, Rood_etal1993, FusiPecci_etal1996}. More recently, with the discovery of multiple populations often extending down to the MS, which have largely confirmed the expectations based on those earlier HB morphology arguments, HB multimodality is once again being used as a diagnostic of the presence of multiple populations in GCs \citep[e.g.,][]{Alonso-Garcia_etal2012}. Here we briefly show that a GC with multiple populations could have a unimodal HB morphology, even though these populations could have a difference in their initial helium abundance, as in the possible case of 47~Tuc \citep[e.g., ][]{Nataf_etal2011, Milone_etal2012}.

First, Fig. \ref{FigZAHB.MvsT} shows the effective temperature of stars when they arrive at the ZAHB with a given mass ($M_{\rm ZAHB}$) for three metallicities, where there is a range of masses that are concentrated near to the minimum $T_{\rm eff}$ of the ZAHB, and this range of masses is higher for higher metallicities. This figure also shows how the relation between $T_{\rm eff}$ and $M_{\rm ZAHB}$ changes with different progenitor masses (upper panel), and different initial helium abundances (lower panel). As can be seen, a decrease in the progenitor mass induces an increase of the minimum $T_{\rm eff}$ of the ZAHB, the effect being greater for lower metallicities. This decrease also induces a small increase of the $M_{\rm ZAHB}$ for a given $T_{\rm eff}$. When the initial helium abundance is increased the range of masses concentrated near the minimum $T_{\rm eff}$ of the ZAHB also increases, which also induces a decrease of $M_{\rm ZAHB}$ for a given $T_{\rm eff}$.

For estimating the mean $T_{\rm eff}$ of a stellar population one can use the final mass of stars at the RGB tip ($M_{\rm Tip}$) depending on the age for a given metallicity and initial helium abundance. Assuming that the difference between $M_{\rm Tip}$ and $M_{\rm ZAHB}$ is small (not a bad approximation considering the short time spent  in the pre-ZAHB phase), one can estimate $M_{\rm ZAHB}$ knowing the age of the GC in study, and then estimate the mean $T_{\rm eff}$. Figure \ref{FigZAHB.TvsAge} shows the change in $T_{\rm eff}$ depending on the age for two coeval initial helium abundances (Y$=0.245$ and Y$=0.345$) and 3 different metallicities, where continuous lines show the $T_{\rm eff}$ at $M_{\rm ZAHB}=M_{\rm Tip}$, and dotted lines represent the effect of a possible dispersion in the mass loss rate, which show the $T_{\rm eff}$ at $M_{\rm ZAHB}=M_{\rm Tip}\pm 0.02 M_\odot$ \citep{Rood_Crocker1989, Catelan_etal2001, Valcarce_Catelan2008}. The additional dotted-dashed line represents an initial helium abundance of Y$=0.295$. As can be seen, a 13 Gyr old GC with higher metallicity (right panel) can hide populations with a $\Delta Y=0.10$ with the initial helium abundance within a small range of temperatures of the red ZAHB, but {\em these population will have a great difference in their luminosity}. For the same age, metal-intermediate and metal-poor GCs (middle and left panel of Fig. \ref{FigZAHB.TvsAge}, respectively) cannot hide multiple populations with a large difference in the initial helium abundance due to the consequent differences in $T_{\rm eff}$. However, smaller differences in Y can be indistinguishable for lower metallicities. Of course, evolutionary effects also play a r\^ole in the shape of the HB, but for studying these effects HB evolutionary tracks and Monte Carlo simulations are required, which is not the focus of this paper.

Finally, we have to recall that here we are using eq. \ref{SC05} with $\eta=1.0$ to take into account the mass loss during the RGB evolution for all stars, irrespective of its chemical composition. Unfortunately, there are at present no strong constraints available on a possible dependence of $\eta$ on Y. On the other hand, $\dot{M}$ in He-rich populations could in principle be constrained, in view of the strong dependence of HB morphology on the He abundance. Future work along these lines is strongly encouraged.

\section{Conclusions}
\label{conclu}

In this work, we studied the effects that must be observed in the theoretical $\log (L/L_\odot)-\log (T_{\rm eff})$ plane for isochrones and ZAHB loci with different initial helium abundances and fixed metallicity, which is the most common hypothesis to explain the multiple populations observed in the color-magnitude diagrams in some GCs. Our study is based on stellar evolutionary calculations for different evolutionary phases in the life of a low-mass star.

First, we presented the PGPUC SEC that was updated with the most recent physical ingredients available in the literature, including opacities, both radiative and conductive \citep{Iglesias_Rogers1996, Ferguson_etal2005, Cassisi_etal2007}; thermonuclear reaction rates \citep{Angulo_etal1999, Kunz_etal2002, Formicola_etal2004, Imbriani_etal2005}; equation of state \citep{Irwin2007}; mass loss \citep{Schroder_Cuntz2005}; and boundary conditions \citep{Catelan2007}. Even though the neutrino energy loss rate was not updated, we recommend the use of the \citet{Kantor_Gusakov2007} interpolated tables for extreme conditions (high densities), which however are not achieved for the stars in our study.

Comparison between PGPUC SEC and other theoretical databases shows similar results. In fact, PGPUC SEC evolutionary tracks and isochrones are in excellent agreement with those produced using the PADOVA, GARSTEC and (specially) BaSTI codes. 

We create a large set of evolutionary tracks for a wide range of metallicities and helium abundances for a GS98 chemical composition and [$\alpha$/Fe]=+0.3, which are available in the PGPUC Online web page. Then theoretical isochrones and ZAHB loci were used to show the effects of different helium abundances at different evolutionary phases along the HR diagram, where in isochrones:

\begin{itemize}
\item At the MS, the separation induced by Y increases for high Z.

\item At the SGB, the separation induced by Y increases for high Z, but it decreases with the age.

\item At the base of the RGB, the difference in $T_{\rm eff}$ induced by Y increases for high Z and also increase with the age.
\end{itemize}

\noindent while at the ZAHB loci:

\begin{itemize}
\item When Y increases, the $L$ of the cooler HB stars increases, but the $L$ of the hotter HB stars decreases.

\item When the age increases, the $L$ of the hotter HB stars increase due to the increase of the $M_{cHe}$.

\item When Z increases, there is an increase in the separation in $L$ due to a difference in Y at the cooler HB stars, but the luminosity difference induced by the age decreases. 
\end{itemize}

We emphasized the importance of the age in determining the ZAHB locus \citep[as pointed out by][and references in Sect.\ref{ZAHBage}]{Caputo_deglInnocenti1995}, and also the location of stars when they arrive to the ZAHB loci depending on the age. In the next papers of this series we will study these effects in different filters for comparing with GCs data, and we will also create a new set of evolutionary tracks with different CNO ratios.

\subsection*{Acknowledgments}

We thank the anonymous referee for her/his comments, which have helped improve the presentation of our results. A.A.R.V. thanks Santi Casissi for his useful comments during A.A.R.V.'s Ph.D. Thesis on this topic. Support for A.A.R.V. and M.C. is provided by the Ministry for the Economy, Development, and Tourism's Programa Iniciativa Cient\'{i}fica Milenio through grant P07-021-F, awarded to The Milky Way Millennium Nucleus; by Proyecto Basal PFB-06/2007; by FONDAP Centro de Astrof\'{i}sica 15010003; by Proyecto FONDECYT Regular \#1110326; and by Proyecto Anillo de Investigaci\'{o}n en Ciencia y Tecnolog\'{i}a PIA CONICYT-ACT 86. A.A.R.V. acknowledges additional support from CNPq, CAPES, from Proyecto ALMA-Conicyt 31090002, MECESUP2, and SOCHIAS.

\bibliographystyle{aa}
\bibliography{avalcarce}

\appendix
\section{PGPUC Online}
\label{AppPGPUConline}

\begin{figure}
   \centering
      \includegraphics[height=20.5cm, trim=0.7cm 1.2cm 12.5cm 0.7cm]{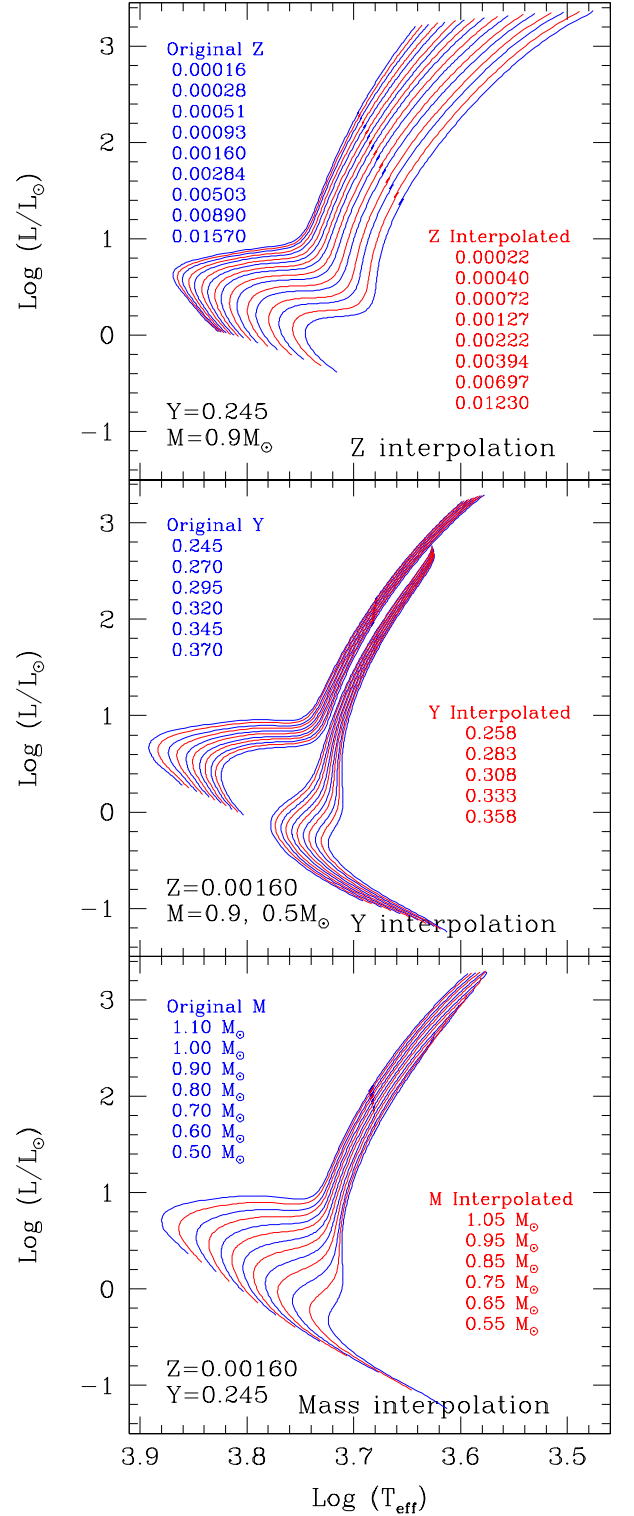}
      \caption{Interpolated evolutionary tracks (red lines) from the original theoretical set of evolutionary tracks (blue lines). Upper panel: metallicity interpolation (metallicity increases from left to right) with constant helium abundance ${\rm Y}=0.245$ and mass of $0.9 M_\odot$. Middle panel: helium interpolation (helium increases from right to left) with constant metallicity ${\rm Z}=0.0016$ and masses of $0.9$ (hotter) and $0.5 M_\odot$ (cooler). Bottom panel: mass interpolation (mass increases from right to left) with constant metallicity ${\rm Z}=0.0016$ and helium abundance ${\rm Y}=0.245$. }
      \label{FigIntepTracks}
 \end{figure}

PGPUC Online (\url{http://www2.astro.puc.cl/pgpuc}) is our web page containing all the PGPUC evolutionary tracks and ZAHB loci. In this web page the Hermite interpolation algorithm for constructing reasonable analytic curves through discrete data points presented by \citet{Hill1982} is used, which produces evolutionary tracks from the MS to the RGB tip for any mass, helium, and metallicities in the ranges $0.5 M_\odot \le M \le 1.1 M_\odot$, $0.230 \le {\rm Y} \le 0.370$, and $0.00016 \le {\rm Z} \le 0.01570$, respectively, for an $[\alpha/{\rm Fe}]=0.3$ and a SC05 mass loss rate with $\eta=1.0$. More especificaly, to interpolate an evolutionary track of a given mass $M_x$, helium abundance ${\rm Y}_x$, and metallicity ${\rm Z}_x$ within a range of masses ($\{M_i\}$, with i=1...7), helium abundances ($\{{\rm Y}_j\}$, with j=1...7), and metallicities ($\{{\rm Z}_k\}$, with k=1...9) we follow the next steps:

\begin{itemize}

\item Using the first EEP of each evolutionary track with mass $M=M_1$, helium abundance ${\rm Y=Y}_1$, and all the 9 metallicities ${\rm Z}_k$, the first EEP of the track with $M=M_1$, ${\rm Y=Y}_1$, and ${\rm Z=Z}_x$ is interpolated. This process is followed for each EEP of the evolutionary track with $M=M_1$ and ${\rm Y=Y}_1$ with the end result that an interpolated evolutionary track with $M=M_1$, ${\rm Y=Y}_1$, and ${\rm Z=Z}_x$ is created.

\item Following the previous process for each of the seven masses $M_i$, we have an interpolated set of evolutionary tracks with seven masses $M_i$, a helium abundance ${\rm Y=Y}_1$, and a metallicity ${\rm Z=Z}_x$.

\item Doing the previous two steps for each helium abundance ${\rm Y}_j$, one has a set of evolutionary tracks with seven masses $M_i$, for each of the seven helium abundances ${\rm Y}_j$, and with a given ${\rm Z=Z}_x$.

\item Now, using the first EEP of each evolutionary track with mass $M=M_1$, metallicity ${\rm Z}_x$, and all the 7 helium abundances ${\rm Y}_j$, the first EEP of the track with ${\rm Y=Y}_x$, that obviously has a ${\rm Z=Z}_x$, is interpolated. This process is followed for each EEP of the evolutionary track with $M=M_1$ and ${\rm Z=Z}_x$, with the end result that an interpolated evolutionary track with $M=M_1$, ${\rm Y=Y}_x$, and ${\rm Z=Z}_x$ is created.

\item When the previous step is repeated for each mass $M_i$ a set of evolutionary tracks with 7 masses $M_i$, ${\rm Y=Y}_x$, and ${\rm Z=Z}_x$ is obtained.

\item Finally, using the first EEP of each mass $M_i$ one can interpolate the first EEP of the mass $M_x$. Repeating this process for each EEP, one creates the final interpolated evolutionary track with $M=M_x$, ${\rm Y=Y}_x$, and ${\rm Z=Z}_x$ (see Fig. \ref{FigIntepTracks}).

\end{itemize}

\begin{figure}
   \centering
      \includegraphics[height=20.5cm, trim=0.7cm 1.2cm 12.5cm 0.7cm]{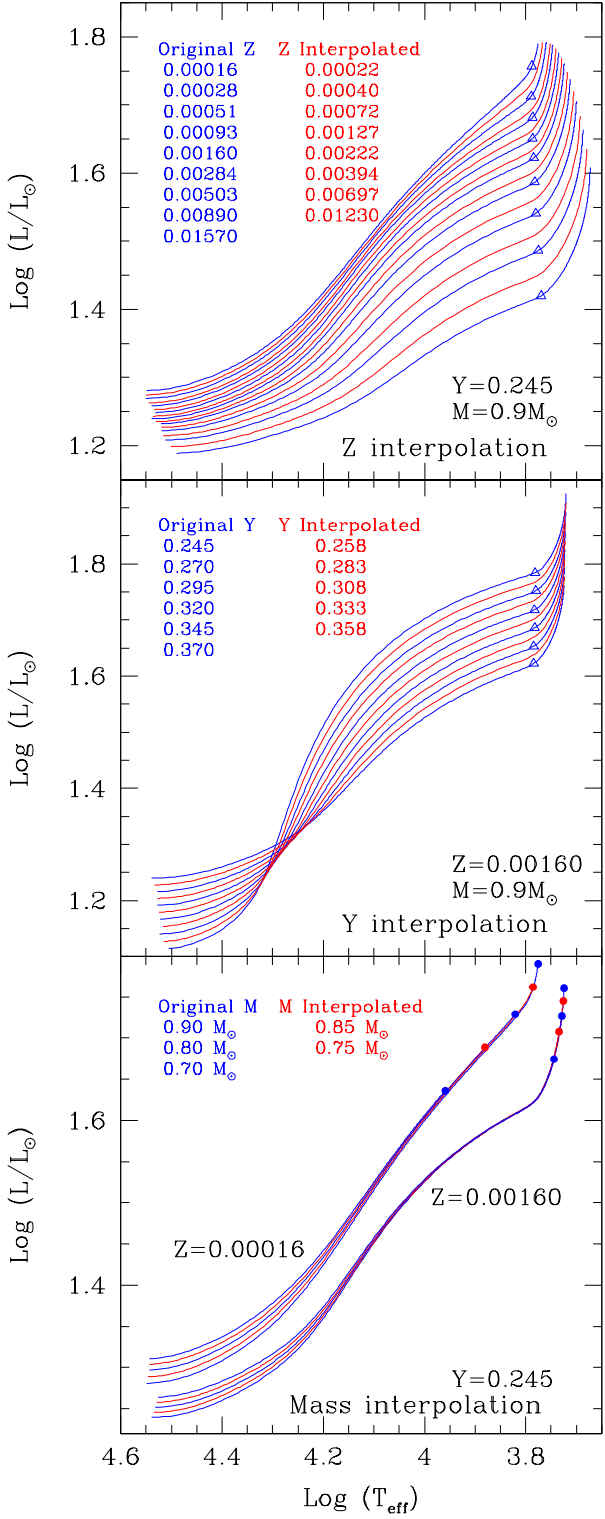}
      \caption{Interpolated ZAHB loci (red lines) from the original theoretical set of ZAHB loci (blue lines). Open triangles show the locus where the convective envelope dissapears. Upper panel: Z interpolation (Z increases from top to bottom) with constant ${\rm Y}=0.245$ and $M_i=0.9 M_\odot$. Middle panel: Y interpolation (Y increases from bottom to top at the red ZAHB and inversely at the blue ZAHB) with constant ${\rm Z}=0.0016$ and $M_i=0.9 M_\odot$. Bottom panel: $M_i$ ($M_i$ increases from bottom to top at the blue ZAHB) with two constant Z (${\rm Z}=0.0016$ and ${\rm Z}=0.00016$) and ${\rm Y}=0.245$. Dots are the initial points for each ZAHB locus, with the mass increasing from bottom to top.}
      \label{FigIntepZAHBs}
 \end{figure}

In the case that we have the same set of evolutionary tracks, but for different $\alpha$-element enhancements $[\alpha /{\rm Fe}]_l$, one can follow the previous procedure for each $[\alpha /{\rm Fe}]_l$, ending with the interpolation of the EEPs for each evolutionary track with $M=M_x$, ${\rm Y=Y}_x$, ${\rm Z=Z}_x$, and $[\alpha /{\rm Fe}]_l$ to create ones with a given $\alpha$-element enhancement. Of course, this procedure can be done for each initial property that can be parameterized (e.g., mass loss, mixing length parameter, overshooting, C+N+O abundance, etc.), which is our intention with the PGPUC online web page: that the user can select the initial property that he or she wants.

Moreover, since any evolutionary track can be created, in this web page it is also possible to create any isochrones within the previous ranges of helium and metallicity. However, due to the fact that the maximum mass for each chemical composition is $1.1 M_\odot$, the minimum age limit is not constant, with a value equal to the required time for the star with $1.1 M_\odot$ of the given chemical composition to evolve from the MS to the RGB tip. In the near future, we plan to increase the maximum mass value.

In the case of the ZAHB, they can also be interpolated for any chemical composition within the range of chemical compositions. This is done using the so-called equivalent ZAHB points (EZAHBPs), which are defined using two criteria, depending on whether the ZAHB star has a convective envelope or not. In Figure \ref{FigIntepZAHBs} the open triangles show the points which divide stars with and without a convective envelope, cooler stars being the ones with convective envelopes. For stars with a convective envelope, 34 EZAHBPs are defined depending on the percentage of convective envelope mass, from the coolest point of the ZAHB (the star with higher convective envelope mass) until the point without a convective envelope. For ZAHB stars without a convective envelope, 100 EZAHBPs are defined, depending on the percentage of envelope mass that ZAHB stars have, from the point without a convective envelope (100\% of envelope mass) until the hottest point at the given ZAHB (0\% of envelope mass). Using the EZAHBPs it is possible to interpolate directly in metallicity and helium abundance (see Fig. \ref{FigIntepZAHBs}). However, the method used to interpolate between progenitor masses is more complex. First, the EZAHBPs can be used only from the hottest ZAHB point until the coolest point of the minimum progenitor mass (in this case, $0.7 M_\odot$). The cool part of the ZAHB is extrapolated from the ZAHB with the nearest higher progenitor mass weighted according to the difference in the stellar parameters between the three progenitor masses. As can be seen from Fig. \ref{FigIntepZAHBs}, we note that the interpolated ZAHB loci are quite satisfactory.

\end{document}